\documentclass[traditabstract]{aa} 
\usepackage{graphicx,color}
\begin{document}
   \title{Kinetic theory  of spatially inhomogeneous stellar systems \\
   without collective effects}


   \author{Pierre-Henri Chavanis
          }

   \institute{Laboratoire de Physique Th\'eorique (IRSAMC), CNRS and UPS, Universit\'e de Toulouse, France
             }



\abstract {We review and complete the kinetic theory of spatially inhomogeneous stellar systems when collective effects (dressing of the stars by their polarization cloud) are neglected. We start from the BBGKY hierarchy issued from the Liouville equation and consider an expansion in powers of $1/N$ in a proper thermodynamic limit. For $N\rightarrow +\infty$, we obtain the Vlasov equation describing the evolution of collisionless stellar systems like elliptical galaxies. At the order $1/N$, we obtain a kinetic equation describing the evolution of collisional stellar systems like globular clusters. This equation coincides with the generalized Landau equation derived by Kandrup (1981) from a more abstract projection operator formalism. This equation does not suffer logarithmic divergences at large scales since spatial inhomogeneity is explicitly taken into account. Making a local approximation, and  introducing an upper  cut-off at the Jeans length, it reduces to the Vlasov-Landau equation which is the standard kinetic equation of stellar systems. Our approach provides a simple and pedagogical derivation of these important equations from the BBGKY hierarchy which is more rigorous for systems with long-range interactions than the two-body encounters theory. Making an adiabatic approximation, we write the generalized Landau equation in angle-action variables and obtain a Landau-type kinetic equation that is valid for fully inhomogeneous stellar systems and is free of divergences at large scales. This equation is less general than the Lenard-Balescu-type kinetic equation recently derived by Heyvaerts (2010) since it neglects collective effects, but it is substantially simpler and could be useful as a first step.
We discuss the evolution of the system as a whole and the relaxation
of a test star in a bath of field stars. We derive the corresponding
Fokker-Planck equation in angle-action variables and provide
expressions for the diffusion coefficient and friction force.

}


\titlerunning{Kinetic theory  of spatially inhomogeneous stellar systems
   without collective effects}

   \maketitle
%

\section{Introduction}

In its simplest description, a stellar system can be viewed as a collection of
$N$ classical point mass stars in Newtonian gravitational interaction (Binney and Tremaine 2008).
As understood early by H\'enon (1964), self-gravitating systems experience
two successive types of relaxation: A rapid ``collisionless'' relaxation towards a
quasi stationary state (QSS) that is a virialized state in mechanical equilibrium
but not in thermodynamical equilibrium, followed by a slow ``collisional''
relaxation. One might think that, due to the development of stellar encounters, the system will reach, for sufficiently large times, a statistical equilibrium state described by the Maxwell-Boltzmann distribution. However, it is well-known that  unbounded stellar systems cannot be in strict statistical equilibrium\footnote{For reviews about the statistical mechanics of self-gravitating systems see, e.g., Padmanabhan (1990), Katz (2003), and Chavanis (2006).} due to the permanent escape of high energy stars (evaporation) and the gravothermal catastrophe (core collapse). Therefore, the statistical mechanics of stellar systems is essentially an out-of-equilibrium problem which must be approached through kinetic theories.

The first kinetic equation was written by Jeans (1915). Neglecting encounters between stars, he
described the dynamical evolution of stellar systems by the
collisionless Boltzmann equation coupled to the Poisson equation. This purely mean field description applies to large groups of stars such as elliptical galaxies whose ages are much less than the collisional relaxation time. A
similar equation was introduced by Vlasov (1938) in plasma physics to describe the collisionless evolution of a system of electric charges interacting by the Coulomb force. The collisionless Boltzmann equation coupled self-consistently to the Poisson equation is oftentimes called the Vlasov equation\footnote{See H\'enon (1982) for a discussion about  the name that one should give to that equation.}, or the Vlasov-Poisson system.

The concept of collisionless relaxation was first
understood by H\'enon (1964) and King (1966). Lynden-Bell (1967) developed a
statistical theory of this process and coined the term ``violent relaxation''. In
the collisionless regime, the system is described by the
Vlasov-Poisson system. Starting from an unsteady or unstable initial condition, the Vlasov-Poisson system develops a complicated mixing process in phase space. The Vlasov equation, being time-reversible, never achieves a steady state but develops filaments at smaller and smaller scales. However, the coarse-grained distribution function usually achieves a steady state on a few dynamical times. Lynden-Bell (1967) tried to predict the QSS resulting from violent relaxation by developing a statistical mechanics of the Vlasov equation. He derived a distribution function formally equivalent to the Fermi-Dirac distribution (or to a superposition of Fermi-Dirac distributions). However, when coupled to the Poisson equation, these distributions have an infinite mass. Therefore, the Vlasov-Poisson system has no statistical equilibrium state (in the sense of Lynden-Bell). This is a clear evidence that violent relaxation is incomplete (Lynden-Bell 1967). Incomplete relaxation is due to inefficient mixing and non-ergodicity. In general, the fluctuations of the gravitational potential $\delta\Phi({\bf r},t)$ that drive the collisionless relaxation last only for a few dynamical times and die out before the system has mixed efficiently (Tremaine et al. 1986). Understanding the origin of incomplete relaxation, and developing models of incomplete violent relaxation to predict the structure of galaxies, is a very difficult problem (Arad \& Johansson 2005). Some models of incomplete violent relaxation have been proposed based on different physical arguments (Bertin and Stiavelli 1984, Stiavelli and Bertin 1987, Hjorth and Madsen 1991, Chavanis 1998, Levin et al. 2008).

On longer timescales, stellar encounters
(sometimes referred to as ``collisions'' by an abuse of language) must
be taken into account. This description is particularly important for small groups of stars such as globular clusters whose ages are of the order of the collisional relaxation time.  Chandrasekhar (1942,1943a,1943b) developed a kinetic
theory of stellar systems in order to determine the timescale of
collisional relaxation and the rate of escape of stars from globular
clusters\footnote{Early estimates of the relaxation time of stellar systems were made by Schwarzschild (1924), Rosseland (1928), Jeans (1929), Smart (1938), and Spitzer (1940). On the other hand, the evaporation time was first estimated by Ambartsumian (1938) and Spitzer (1940).}. To simplify the kinetic theory, he considered an infinite
homogeneous system of stars.
He started from the general Fokker-Planck equation
and determined the diffusion coefficient and the friction force (first
and second moments of the velocity increments) by considering the mean
effect of a succession of two-body encounters\footnote{Later, Cohen et al. (1950), Gasiorowicz et al. (1956), and Rosenbluth et al. (1957) proposed a simplified derivation
of the coefficients of diffusion and
friction.}. However, his
approach leads to a logarithmic divergence at large scales that is
more difficult to remove in stellar dynamics than in plasma physics
because of the absence of Debye shielding for the gravitational
force\footnote{A few years earlier, Landau (1936) had developed a kinetic theory of Coulombian plasmas based on two-body encounters. Starting from the Boltzmann (1872) equation, and making a weak deflection approximation, he derived a kinetic equation for the collisional evolution of neutral plasmas. His approach leads to a divergence at large scales that he removed heuristically by introducing a cut-off at the Debye length (Debye and H\"uckel 1923) which is the size over which the electric field produced by a charge is screened by the cloud of opposite charges. Later, Lenard (1960) and
Balescu (1960) developed a more precise  kinetic
theory taking collective effects into account. They derived a more elaborate kinetic equation, free of divergence at large scales, in which the Debye length appears naturally. This justifies the heuristic procedure of Landau.}. Chandrasekhar and von Neumann (1942) developed a completely stochastic formalism of gravitational
fluctuations and showed that the fluctuations of the gravitational
force are given by the Holtzmark distribution (a particular L\'evy
law) in which the nearest neighbor plays a prevalent role. From these
results, they argued that the logarithmic divergence has to be cut-off
at the interparticle distance (see also Jeans 1929). However, since the interparticle
distance is smaller than the Debye length, the same arguments should
also apply in plasma physics, which is not the case. Therefore, the
conclusions of Chandrasekhar and von Neumann are usually taken with
circumspection. In particular, Cohen et al. (1950)  argue
that the logarithmic divergence should be cut-off at the Jeans length
which gives an estimate of the system's size. Indeed, while in neutral plasmas the effective interaction
distance is limited to the Debye length, in a self-gravitating system,
the distance between interacting particles is only limited by the
system's size. Therefore, the
Jeans length is the gravitational analogue of the Debye
length. These kinetic theories lead to a collisional relaxation
time scaling as $t_{R}\sim (N/\ln N) t_D$, where $t_D$ is the dynamical time
and $N$ the number of stars in the system.  Chandrasekhar (1949) also developed a Brownian theory of stellar dynamics and
showed that, from a qualitative point of view, the results of kinetic
theory can be understood very simply in that framework. In particular,
he showed that a dynamical friction is necessary to maintain the
Maxwell-Boltzmann distribution of statistical equilibrium and that the
coefficients of friction and diffusion are related to each other by an
Einstein relation which is a manifestation of the fluctuation-dissipation theorem. This relation is
confirmed by his more precise kinetic theory based on two-body
encounters.   It is important to emphasize, however, that Chandrasekhar
did not derive the kinetic equation for the evolution of the system as
a whole. Indeed, he considered the Brownian motion of a test star in a
{\it fixed} distribution of field stars (bath) and derived the corresponding
Fokker-Planck equation. This equation has been used
by Chandrasekhar (1943b), Spitzer and H\"arm (1958), Michie (1963), King (1965),
and more recently Lemou and Chavanis (2010) to study the evaporation of stars from globular
clusters in a simple setting.

King (1960) noted that, if we were to describe the
dynamical evolution of the cluster as a whole, the distribution of the
field stars must evolve in time in a self-consistent manner so
that the kinetic equation must be an integrodifferential equation. The
kinetic equation obtained by King, from the results of Rosenbluth et al. (1957),
is equivalent to the Landau equation, although written in a different form.
It is interesting to note, for historical reasons, that none of the previous authors
seemed to be aware of the work of Landau (1936)  in plasma physics. There is, however, an important difference between stellar
dynamics and plasma physics. Neutral plasmas are spatially homogeneous
due to Debye shielding.  By contrast, stellar systems are spatially
inhomogeneous. The above-mentioned kinetic theories developed for an
infinite homogeneous system can be applied to an inhomogeneous system
only if we make a {\it local approximation}. In that case, the
collision term is calculated as if the system were spatially
homogeneous or as if the collisions could be treated as local. Then,
the effect of spatial inhomogeneity is only retained in the advection
(Vlasov) term which describes the evolution of the system due to
mean-field effects. This leads to the
Vlasov-Landau equation which is the standard kinetic equation of
stellar dynamics. To our knowledge, this equation
has been first written  (in a different form), and studied, by H\'enon
(1961). H\'enon also exploited the timescale separation between
the dynamical time $t_D$ and the relaxation time $t_{R}\gg t_D$ to
derive a simplified kinetic equation for $f(\epsilon,t)$, where
$\epsilon=v^2/2+\Phi({\bf r},t)$ is the individual energy of a star by
unit of mass, called the orbit-averaged Fokker-Planck equation. In
this approach, the distribution function $f({\bf r},{\bf v},t)$,
averaged over a short timescale, is a steady state of the Vlasov
equation of the form $f(\epsilon,t)$ which slowly evolves in time, on
a long timescale, due to the development of ``collisions''
(i.e. correlations caused by finite $N$ effects or graininess). H\'enon 
used this equation to obtain a more relevant value for 
the rate of evaporation from globular clusters, valid for inhomogeneous systems. Cohn
(1980) numerically solved the orbit-averaged Fokker-Planck
equation to describe the collisional evolution of star clusters.  His treatment
accounts both for the escape of high energy stars put forward by
Spitzer (1940) and for the phenomenon of core collapse
resulting from the gravothermal catastrophe discovered by Antonov
(1962)  and Lynden-Bell and Wood (1968) on the basis of thermodynamics and
statistical mechanics.  The local approximation, which is a crucial
step in the kinetic theory, is supported by the stochastic approach of
Chandrasekhar and von Neumann (1942) showing the preponderance of the nearest
neighbor. However, this remains a simplifying assumption which is not
easily controllable. In particular, as we have already indicated, the
local approximation leads to a logarithmic divergence at large scales
that is difficult to remove. This divergence would not have occurred
if full account of spatial inhomogeneity had been given since the
start.

The effect of spatial inhomogeneity was investigated by Severne
and Haggerty (1976), Parisot and Severne (1979), Kandrup
(1981), and Chavanis (2008). In particular, Kandrup (1981) derived
a generalized Landau equation from the Liouville equation by using
projection operator technics. This generalized Landau equation is interesting
because it takes into account effects of spatial inhomogeneity 
which were neglected in previous approaches. Since the finite
extension of the system is properly accounted for, there is no
divergence at large scales\footnote{There remains a logarithmic divergence at small scales due to the neglect of strong collisions. This divergence can be cured heuristically by introducing a cut-off at the Landau length which corresponds to the impact parameter leading
to a deflection at $90^{o}$ (Landau 1936).}. Furthermore, this approach clearly show which
approximations are needed in order to recover the traditional Landau
equation. Unfortunately, the generalized Landau equation remains
extremely complicated for practical applications.

In addition, this equation is still approximate as it neglects
collective effects and considers binary collisions between ``naked''
particles. As in any weakly coupled system, the particles engaged in
collisions are ``dressed'' by the polarization clouds caused by their
own influence on other particles. Collisions between dressed particles
have quantitatively different outcomes than collisions between naked
ones. In the case of plasmas, collective effects are responsible for
Debye shielding and they are accounted for in the Lenard-Balescu
equation. They allow to eliminate the logarithmic divergence that
occurs at large scales in the Landau equation. For self-gravitating
systems, they lead to ``anti-shielding'' and are more difficult to
analyze because the system is spatially inhomogeneous\footnote{In a
plasma, since the Coulomb force between electrons is repulsive, each
particle of the plasma tends to attract to it particles of opposite
charge and to repel particles of like charge, thereby creating a kind
of ``cloud'' of opposite charge which screens the interaction at the
scale of the Debye length. In the gravitational case, since the Newton
force is attractive, the situation is considerably different. The test
star draws neighboring stars into its vicinity and these add their
gravitational force to that of the test star itself. The ``bare''
gravitational force of the test star is thus augmented rather than
shielded. The polarization acts to increase the effective
gravitational mass of a test star.}. If we consider a finite
homogeneous system, and take collective effects into account, one
finds a severe divergence of the diffusion coefficient when the size
of the domain reaches the Jeans scale (Weinberg 1993). This
divergence, which is related to the Jeans instability, does
not occur in a stable spatially inhomogeneous stellar system. Some authors
like Thorne (1968), Miller (1968), Gilbert (1968,1970), and Lerche
(1971) attempted to take collective effects and spatial inhomogeneity
into account. However, they obtained very complicated kinetic
equations that have not found application until now. They managed,
however, to show that collective effects are equivalent to increasing
the effective mass of the stars, hence diminishing the relaxation
time. Since, on the other hand, the effect of spatial inhomogeneity is
to increase the relaxation time (Parisot and Severne 1979), the two
effects act in opposite direction and may balance each other.

Recently, Heyvaerts (2010) derived from the BBGKY hierarchy issued
from the Liouville equation a kinetic equation in angle-action
variables that takes both spatial inhomogeneities and collective
effects into account. To calculate the collective response, he used
Fourier-Laplace transforms and introduced a bi-orthogonal basis of
pairs of density-potential functions (Kalnajs 1971a). The kinetic
equation derived by Heyvaerts is the counterpart for spatially
inhomogeneous self-gravitating systems of the Lenard-Balescu equation
for plasmas. Following his work, we showed that this equation could be
obtained equivalently from the Klimontovich equation by
making the so-called quasilinear approximation (Chavanis 2012). We
also developed a test particle approach and derived the corresponding
Fokker-Planck equation in angle-action variables, taking collective
effects into account. This provides general expressions of the
diffusion coefficient and friction force for spatially inhomogeneous
stellar systems.

In a sense, these equations solve the problem of the kinetic theory of
stellar systems since they take into account both spatial
inhomogeneity and collective effects. However, their drawback is that
they are extremely complicated to solve (in addition of being
complicated to derive). In an attempt to reduce the complexity of the
problem, we shall derive in this paper a kinetic equation that is
valid for spatially inhomogeneous stellar systems but that neglects
collective effects. Collective effects may be less crucial in stellar
dynamics than in plasma physics. In plasma physics, they must be taken
into account in order to remove the divergence at large scales that
appears in the Landau equation. In the case of stellar systems, this
divergence is removed by the spatial inhomogeneity of the system, not
by collective effects. Actually, previous kinetic equations based on
the local approximation ignore collective effects and already give
satisfactory results. We shall therefore ignore collective effects and
derive a kinetic equation (in position-velocity and angle-action
variables) that is the counterpart for spatially inhomogeneous
self-gravitating systems of the Landau equation for plasmas. Our
approach has three main interests. First, the derivation of this
Landau-type kinetic equation is considerably simpler than the
derivation of the Lenard-Balescu-type kinetic equation, and it can be
done in the physical space without having to introduce Laplace-Fourier
transforms nor bi-orthogonal basis of pairs of density-potential
functions. This offers a more physical derivation of kinetic equations
of stellar systems that can be of interest for
astrophysicists. Secondly, our approach is sufficient to remove the
large-scale divergence that occurs in kinetic theories based on the
local approximation. It represents therefore a conceptual progress in
the kinetic theory of stellar systems. Finally, this equation is
simpler than the Lenard-Balescu-type kinetic equation derived by
Heyvaerts (2010), and it could be useful in a first step before
considering more complicated effects. Its drawback is to ignore
collective effects but this may not be crucial as we have explained
(as suggested by Weinberg's work, collective effects become important
only when the system is close to instability). This Landau-type
equation was previously derived for systems with arbitrary long-range
interactions\footnote{For a review on the dynamics and thermodynamics
of systems with long-range interactions, see Campa et al. (2009).} in
various dimensions of space (see Chavanis 2007,2008,2010) but we think
that it is important to discuss these results in the specific case of
self-gravitating systems with complements and amplification.

The paper is organized as follows. In Section \ref{sec_whole}, we study the dynamical evolution of a spatially inhomogeneous stellar system as a whole. Starting from the BBGKY hierarchy issued from the Liouville equation, and neglecting collective effects, we derive a general kinetic equation valid at the order $1/N$ in a proper
thermodynamic limit. For $N\rightarrow +\infty$, it reduces to the Vlasov equation. At the order $1/N$ we recover the generalized Landau equation derived by Kandrup (1981) from a more abstract projection operator formalism. This equation is free of divergence at large scales since spatial inhomogeneity has been properly accounted for. Making a local approximation and introducing an upper cut-off at the Jeans length, we recover the standard Vlasov-Landau equation which is usually derived from a kinetic theory based on two-body encounters. Our approach provides an alternative derivation of this fundamental equation from the more rigorous Liouville equation. It has therefore some pedagogical interest. In Section \ref{sec_tp}, we study the relaxation of a test star in a steady distribution of field stars. We derive the corresponding Fokker-Planck equation and determine the expressions of the diffusion and friction coefficients. We emphasize the difference between the friction by polarization and the total friction (this difference may have been overlooked in previous works). For a thermal bath, we derive the Einstein relation between the diffusion and friction coefficients and obtain the explicit expression of the diffusion tensor. This returns the standard results obtained from the two-body encounters theory but, again, our presentation is different and offers an alternative  derivation of these important results. For that reason, we give a short review of the basic formulae. 
 In Section \ref{sec_angleaction}, we derive a Landau-type kinetic equation written in angle-action variables and discuss its main properties. This equation, which does not make the local approximation, applies to fully inhomogeneous stellar systems and is free of divergence at large scales. We also develop a test particle approach and derive the corresponding Fokker-Planck equation in angle-action variables. Explicit expressions are given for the diffusion tensor and friction force, and they are compared with previous expressions obtained in the literature.

\section{Evolution of the system as a whole}
\label{sec_whole}

\subsection{The BBGKY hierarchy}
\label{sec_bbgky}

We consider an isolated system of $N$ stars with identical mass $m$ in
Newtonian interaction. Their dynamics is fully described by the
Hamilton equations
\begin{eqnarray}
\label{bbgky1} m{d{\bf r}_{i}\over dt}={\partial H\over\partial {\bf v}_{i}}, \qquad m{d{\bf v}_{i}\over dt}=-{\partial H\over\partial {\bf r}_{i}},\nonumber\\
H={1\over 2}\sum_{i=1}^{N}m{v_{i}^{2}}-G m^{2}\sum_{i<j} \frac{1}{|{\bf r}_{i}-{\bf r}_{j}|}.
\end{eqnarray}
This Hamiltonian system conserves the energy $E=H$, the mass $M=Nm$,
and the angular momentum ${\bf L}=\sum_{i} m {\bf r}_{i}\times {\bf
v}_{i}$. As recalled in the Introduction, stellar systems cannot reach a statistical equilibrium state in a strict sense. In order to understand their evolution, it is necessary to develop a kinetic theory.

We introduce the $N$-body distribution function
$P_N({\bf r}_{1},{\bf v}_{1},...,{\bf r}_{N},{\bf v}_{N},t)$ giving
the probability density of finding, at time $t$, the first star with position
${\bf r}_1$ and velocity ${\bf v}_1$, the second star with position
${\bf r}_2$ and velocity ${\bf v}_2$ etc. Basically, the evolution of the $N$-body distribution function is
governed by the Liouville equation
\begin{equation}
\label{bbgky2} {\partial P_{N}\over\partial t}+\sum_{i=1}^{N}\biggl
({\bf v}_{i}\cdot {\partial P_{N}\over\partial {\bf r}_{i}}+{\bf
F}_{i}\cdot {\partial P_{N}\over\partial {\bf v}_{i}}\biggr
)=0,
\end{equation}
where
\begin{equation}
\label{bbgky3}
{\bf F}_{i}=-{\partial \Phi_d\over\partial {\bf r}_{i}}=-G m\sum_{j\neq i}\frac{{\bf r}_{i}-{\bf r}_{j}}{|{\bf r}_{i}-{\bf r}_{j}|^{3}}=\sum_{j\neq i}{\bf F}(j\rightarrow i),
\end{equation}
is the gravitational force by unit of mass experienced by the $i$-th
star due to its interaction with the other stars. Here, $\Phi_d({\bf r})$ denotes
the exact gravitational potential produced by the discrete
distribution of stars and ${\bf F}(j\rightarrow i)$ denotes the exact
force by unit of mass created by the $j$-th star on the $i$-th star.
The Liouville equation (\ref{bbgky2}),
which is equivalent to the Hamilton equations (\ref{bbgky1}), contains
too much information to be exploitable. In practice, we are only
interested in the evolution of the one-body distribution $P_1({\bf r},{\bf v},t)$.

From the
Liouville equation (\ref{bbgky2}) we can construct the complete BBGKY
hierarchy for the reduced distribution functions
\begin{equation}
\label{bbgky4}
P_{j}({\bf x}_{1},...,{\bf x}_{j},t)=\int P_{N}({\bf x}_{1},...,{\bf x}_{N},t)\, d{\bf x}_{j+1}...d{\bf x}_{N},
\end{equation}
where the notation ${\bf x}$ stands for $({\bf r},{\bf v})$. The generic term of this hierarchy
reads
\begin{eqnarray}
\label{bbgky5} {\partial P_{j}\over\partial t}+\sum_{i=1}^{j}{\bf v}_{i}\cdot {\partial
P_{j}\over\partial {\bf r}_{i}}+\sum_{i=1}^{j}\sum_{k=1,k\neq i}^{j} {\bf F}(k\rightarrow i)\cdot {\partial P_{j}\over \partial {\bf v}_{i}}\nonumber\\
+(N-j)\sum_{i=1}^{j}\int {\bf F}(j+1\rightarrow i)\cdot {\partial P_{j+1}\over\partial {\bf v}_{i}}\, d{\bf x}_{j+1}=0.
\end{eqnarray}
This hierarchy of equations is not closed since the equation for the
one-body distribution $P_{1}({\bf x}_{1},t)$ involves the two-body
distribution $P_{2}({\bf x}_{1},{\bf x}_{2},t)$, the equation for the
two-body distribution $P_{2}({\bf x}_{1},{\bf x}_{2},t)$ involves the
three-body distribution $P_{3}({\bf x}_{1},{\bf x}_{2},{\bf
x}_{3},t)$, and so on.

It is convenient to introduce a cluster representation of the distribution functions. Specifically, we can express the reduced distribution $P_{j}({\bf x}_{1},...,{\bf x}_{j},t)$ in terms of products of distribution functions $P_{j'<j}({\bf x}_{1},...,{\bf x}_{j'},t)$ of lower order plus a correlation function $P_{j}'({\bf x}_{1},...,{\bf x}_{j},t)$ [see, e.g.,  Eqs. (\ref{trunc3}) and (\ref{trunc4}) below]. Considering the scaling of the terms in each equation of the BBGKY
hierarchy, we can see that there exist solutions of the whole BBGKY hierarchy such that the
correlation functions $P_{j}'$ scale as $1/N^{j-1}$ in the proper thermodynamic limit $N\rightarrow +\infty$ defined in Appendix \ref{sec_tl}. This implicitly assumes that the initial condition has
no correlation, or that the initial correlations respect this scaling\footnote{If there are non-trivial correlations in the initial state, like primordial binary
stars, the kinetic theory will be different from the one developed
in the sequel.}.  If this scaling is satisfied, we can consider an expansion of 
the BBGKY hierarchy in terms of the small parameter $1/N$. This is similar to the expansion of the BBGKY hierarchy is plasma physics in terms of the small parameter $1/\Lambda$ where $\Lambda\gg 1$ represents the number of charges in the Debye sphere (Balescu 2000). However, in
plasma physics, the system is spatially homogeneous (due to Debye
shielding which restricts the range of interaction) while, for stellar systems, spatial inhomogeneity must be taken into account. This brings additional terms in the BBGKY hierarchy that are absent in
plasma physics.

\subsection{The truncation of the BBGKY hierarchy at the order $1/N$}
\label{sec_trunc}

The first two equations of the BBGKY hierarchy are
\begin{eqnarray}
\frac{\partial P_1}{\partial t}(1)&+&{\bf v}_1\cdot \frac{\partial P_1}{\partial {\bf r}_1}(1)\nonumber\\
&+&(N-1)\int {\bf F}(2\rightarrow
1)\cdot \frac{\partial P_{2}}{\partial {\bf v}_1}(1,2)\, d{\bf x}_{2}=0,
 \label{trunc1}
\end{eqnarray}
\begin{eqnarray}
\frac{1}{2}\frac{\partial P_2}{\partial t}(1,2)+{\bf v}_1\cdot \frac{\partial P_2}{\partial {\bf r}_1}(1,2)+{\bf F}(2\rightarrow 1)\cdot \frac{\partial P_2}{\partial {\bf
v}_1}(1,2)\nonumber\\
+(N-2)\int {\bf F}(3\rightarrow
1)\cdot \frac{\partial P_{3}}{\partial {\bf v}_1}(1,2,3)\, d{\bf x}_{3}+(1\leftrightarrow 2)=0.
 \label{trunc2}
\end{eqnarray}
We decompose the two- and three-body distributions in the form
\begin{equation}
\label{trunc3}
P_{2}({\bf x}_{1},{\bf x}_{2})=P_{1}({\bf x}_{1})P_{1}({\bf x}_{2})+P_{2}'({\bf x}_{1},{\bf x}_{2}),
\end{equation}
\begin{eqnarray}
\label{trunc4}
P_{3}({\bf x}_{1},{\bf x}_{2},{\bf x}_{3})=P_{1}({\bf x}_{1})P_{1}({\bf x}_{2})P_{1}({\bf x}_{3})+P_{2}'({\bf x}_{1},{\bf x}_{2})P_{1}({\bf x}_{3})\nonumber\\
+P_{2}'({\bf x}_{1},{\bf x}_{3})P_{1}({\bf x}_{2})+P_{2}'({\bf x}_{2},{\bf x}_{3})P_{1}({\bf x}_{1})+P_{3}'({\bf x}_{1},{\bf x}_{2},{\bf x}_{3}),\nonumber\\
\end{eqnarray}
where $P'_j({\bf x}_1,...,{\bf x}_j,t)$ is the correlation function of order $j$. Substituting Eqs. (\ref{trunc3}) and (\ref{trunc4}) in Eqs. (\ref{trunc1}) and (\ref{trunc2}), and simplifying some terms, we obtain
\begin{eqnarray}
\frac{\partial P_1}{\partial t}(1)&+&{\bf v}_1\cdot \frac{\partial P_1}{\partial {\bf r}_1}(1)\nonumber\\
&+&(N-1)\left\lbrack \int {\bf F}(2\rightarrow
1)P_{1}(2)\, d{\bf x}_{2}\right\rbrack\cdot \frac{\partial P_1}{\partial {\bf v}_{1}}(1) \nonumber\\ &=&-(N-1)\frac{\partial}{\partial {\bf v}_1}\cdot \int {\bf F}(2\rightarrow
1) P'_{2}(1,2)\, d{\bf x}_{2},
 \label{trunc5}
\end{eqnarray}
\begin{eqnarray}
\frac{1}{2}\frac{\partial P'_2}{\partial t}(1,2)+{\bf v}_1\cdot \frac{\partial P'_2}{\partial {\bf r}_1}(1,2)+{\bf F}(2\rightarrow 1)\cdot \frac{\partial P'_2}{\partial {\bf
v}_1}(1,2)\nonumber\\
+(N-2) \left\lbrack \int {\bf F}(3\rightarrow
1) P_1(3)\, d{\bf x}_{3} \right\rbrack \cdot \frac{\partial P'_{2}}{\partial {\bf v}_1}(1,2)\nonumber\\
+\left\lbrack {\bf F}(2\rightarrow 1)- \int {\bf F}(3\rightarrow
1) P_1(3)\, d{\bf x}_{3}\right\rbrack \cdot \frac{\partial P_1}{\partial {\bf
v}_1}(1)P_1(2)\nonumber\\
+(N-2)\left\lbrack \int {\bf F}(3\rightarrow
1)  P'_{2}(2,3)\, d{\bf x}_{3}\right\rbrack \cdot \frac{\partial P_1}{\partial {\bf v}_1}(1)
\nonumber\\
-\frac{\partial}{\partial {\bf v}_1}\cdot \int {\bf F}(3\rightarrow
1)  P'_{2}(1,3)P_1(2)\, d{\bf x}_{3}\nonumber\\
+(N-2)\frac{\partial}{\partial {\bf v}_1}\cdot \int {\bf F}(3\rightarrow
1)  P'_{3}(1,2,3)\, d{\bf x}_{3}+(1\leftrightarrow 2)=0.\nonumber\\
 \label{trunc6}
\end{eqnarray}
Equations (\ref{trunc5}) and (\ref{trunc6}) are exact for all $N$ but they are not closed. As explained previously, we shall close these equations at the order $1/N$ in the thermodynamic limit $N\rightarrow +\infty$. In this limit $P'_2\sim 1/N$ and $P'_3\sim 1/N^2$. On the other hand, $P_1\sim 1$ and $|{\bf F}(i\rightarrow j)|\sim G\sim 1/N$ (see Appendix \ref{sec_tl}).

The term in the l.h.s. of Eq. (\ref{trunc5}) is of order $1$, and the term in the r.h.s. is of order $1/N$.  Let us now consider the terms in Eq. (\ref{trunc6}) one by one. The first four terms correspond to the Liouville equation. The Liouvillian ${\cal L}={\cal L}_0+{\cal L}_{12}+\langle{\cal L}\rangle$ describes the complete two-body problem, including the inertial motion, the interaction between the particles $(1,2)$ and the mean field produced by the other particles.  The terms ${\cal L}_0$ and $\langle{\cal L}\rangle$ are of order $1/N$ while the term ${\cal L}_{12}$ is of order $1/N^2$. Therefore, the interaction term ${\cal L}_{12}$ can {\it a priori}\footnote{Actually, the interaction term becomes large at small scales so its effect is not totally negligible (in other words, the expansion in terms of $1/N$ is not uniformly convergent). In particular, the interaction term ${\cal L}_{12}$ must be taken into account in order to describe strong collisions with small impact parameter that lead to large deflections very different from the mean field trajectory corresponding to ${\cal L}_0+\langle{\cal L}\rangle$. We shall return to this problem in Sec. \ref{sec_gen}.} be neglected in the Liouvillian. This corresponds to the {\it weak coupling approximation} where only the mean field term ${\cal L}_0+\langle{\cal L}\rangle$ is retained. The fifth term in Eq. (\ref{trunc6}) is a source term ${\cal S}$ expressible in terms of the one-body distribution; it is of order $1/N$. If we consider only the mean field Liouvillian ${\cal L}_0+\langle{\cal L}\rangle$ and the source term ${\cal S}$, as we shall do in this paper, we can obtain a kinetic equation for stellar systems that is the counterpart of the Landau equation in plasma physics. The sixth  term is of order $1/N$ and it corresponds to collective effects (i.e. dressing of the particles by the polarization cloud). In plasma physics, this term leads to the Lenard-Balescu equation. It takes into account dynamical screening and regularizes the divergence at large scales that appears in the Landau equation. In the case of stellar systems, there is no large-scale divergence because of the spatial inhomogeneity of the system. Therefore, collective effects are less crucial in the kinetic theory of stellar systems than in plasma physics. However, this term has been properly taken into account by Heyvaerts (2010) who obtained a kinetic equation of stellar systems that is the counterpart of the Lenard-Balescu equation in plasma physics.  The last two terms are of the order $1/N^2$ and they will be neglected. In particular, the three-body correlation function $P_3'$, of order $1/N^2$, can be neglected at the order $1/N$. In this way, the hierarchy of equations is closed and a kinetic equation involving only two-body encounters can be obtained.

If we introduce the notations $f=NmP_{1}$
(distribution function) and $g=N^{2}P_{2}'$ (two-body correlation
function), we get at the order $1/N$:
\begin{eqnarray}
{\partial f\over\partial t}(1)+{\bf v}_1\cdot {\partial f\over\partial {\bf r}_1}(1)+\frac{N-1}{N}\langle {\bf F}\rangle(1)\cdot  {\partial f\over \partial {\bf v}_1}(1)=\nonumber\\
-m {\partial \over\partial {\bf v}_1}\cdot \int {\bf F}(2\rightarrow 1)g(1,2)\, d{\bf x}_{2},
\label{trunc7}
\end{eqnarray}
\begin{eqnarray}
\label{trunc8} \frac{1}{2}{\partial g\over\partial t}(1,2)+{\bf v}_1\cdot {\partial g\over\partial {\bf r}_1}(1,2)+\langle {\bf F}\rangle(1)\cdot  {\partial g\over\partial {\bf v}_1}(1,2)\nonumber\\
+\frac{1}{m^{2}}
 {\tilde{\bf F}}(2\rightarrow 1) \cdot   {\partial f\over\partial {\bf v}_1}(1)f(2)+
 {\bf {F}}(2\rightarrow 1)\cdot  \frac{\partial g}{\partial {\bf v}_1}(1,2)\nonumber\\
+ \left\lbrack \int {\bf F}(3\rightarrow 1)g(2,3) \, d{\bf x}_{3}\right\rbrack\cdot \frac{\partial f}{\partial {\bf v}_1}(1)+(1\leftrightarrow 2)=0.
\end{eqnarray}
We have introduced the  mean force (by unit of mass) created on star $1$  by all the other stars
\begin{eqnarray}
\label{trunc9}
\langle {\bf F}\rangle(1) =\int {\bf F}(2\rightarrow 1)\frac{f(2)}{m}\, d{\bf x}_{2}=-\nabla\Phi(1),
\end{eqnarray}
and  the fluctuating force (by unit of mass) created by star $2$
on star $1$:
\begin{eqnarray}
\label{trunc10}
{\tilde{\bf F}}(2\rightarrow 1)={\bf F}(2\rightarrow
1)-\frac{1}{N}\langle {\bf F}\rangle(1).
\end{eqnarray}
Equations (\ref{trunc7}) and (\ref{trunc8}) are exact at the order $1/N$. They form the right basis to develop the kinetic theory of stellar systems
at this order of approximation. Since the collision term in the r.h.s. of Eq. (\ref{trunc7}) is of order $1/N$, we expect that the relaxation time of stellar systems scales as $\sim Nt_D$ where $t_D$ is the dynamical time. As we shall see, the discussion is more complicated due to the presence of logarithmic corrections in the relaxation time and the absence of strict statistical equilibrium state.

\subsection{The limit $N\rightarrow +\infty$: the Vlasov equation (collisionless regime)}
\label{sec_vlasov}

In the limit $N\rightarrow +\infty$, for a fixed interval of time $[0,T]$ (any), the
correlations between stars can be neglected. Therefore, the mean field approximation becomes exact and the $N$-body distribution function factorizes in $N$ one-body
distribution functions
\begin{eqnarray}
P_{N}({\bf x}_{1},...,{\bf
x}_{N},t)=\prod_{i=1}^{N}P_{1}({\bf x}_{i},t).
\label{vlasov1}
\end{eqnarray}
Substituting the factorization (\ref{vlasov1}) in the Liouville equation (\ref{bbgky2}), and integrating over ${\bf x}_2$, ${\bf x}_3$, ..., ${\bf x}_N$, we find that  the smooth distribution
function $f({\bf r},{\bf v},t)=NmP_{1}({\bf r},{\bf v},t)$ is the solution
of the Vlasov equation
\begin{eqnarray}
{\partial f\over\partial t}+{\bf v}\cdot {\partial f\over\partial
{\bf r}}+\langle {\bf F}\rangle \cdot {\partial f\over \partial
{\bf v}}=0, \nonumber\\
\langle {\bf F}\rangle=-\nabla\Phi,\qquad \Delta\Phi=4\pi G\int f \, d{\bf v}. \label{vlasov2}
\end{eqnarray}
This equation also results from Eq. (\ref{trunc7}) if we neglect the
correlation function $g(1,2)$ in the r.h.s. and replace $N-1$ by $N$.

The Vlasov equation describes the {\it collisionless evolution} of stellar
systems for times shorter than the relaxation time $\sim N t_D$.
In practice $N\gg 1$ so that the domain of validity of the
Vlasov equation is huge (see the end of Sec. \ref{sec_shr}). 
As recalled in the Introduction, the Vlasov-Poisson system  develops
a process of phase mixing and violent relaxation
leading to a quasi-stationary state (QSS) on a very short timescale,
of the order of a few dynamical times $t_{D}$. Elliptical galaxies are in such QSSs.
Lynden-Bell (1967) developed a statistical mechanics
of the Vlasov equation in order to describe this process of violent
relaxation and predict the QSS achieved by the system. Unfortunately, the predictions of his statistical theory are limited by the problem of incomplete relaxation. Kinetic theories of violent relaxation, which may account for incomplete relaxation, have been developed by Kadomtsev and Pogutse (1970), Severne and Luwel (1980), and Chavanis (1998,2008).

\subsection{The order $O(1/N)$: the generalized Landau equation (collisional regime)}
\label{sec_gen}

If we neglect strong collisions and collective effects, the first two equations (\ref{trunc7}) and (\ref{trunc8}) of the BBGKY hierarchy reduce to
\begin{eqnarray}
{\partial f\over\partial t}(0)+{\bf v}\cdot {\partial f\over\partial {\bf r}}(0)&+&\frac{N-1}{N}\langle {\bf F}\rangle(0)\cdot  {\partial f\over \partial {\bf v}}(0)=\nonumber\\
&-&m {\partial \over\partial {\bf v}}\cdot \int {\bf F}(1\rightarrow 0)g(0,1)\, d{\bf x}_{1},
\label{gen1}
\end{eqnarray}
\begin{eqnarray}
\label{gen2} \frac{1}{2}{\partial g\over\partial t}(0,1)&+&{\bf v}\cdot {\partial g\over\partial {\bf r}}(0,1)+\langle {\bf F}\rangle(0)\cdot  {\partial g\over\partial {\bf v}}(0,1)+(0\leftrightarrow 1)\nonumber\\
=&-&\frac{1}{m^{2}}
 \tilde{\bf F}(1\rightarrow 0)\cdot  {\partial f\over\partial {\bf v}}(0) f(1)+(0\leftrightarrow 1).
\end{eqnarray}
The first equation gives the evolution of the one-body distribution function. The l.h.s. corresponds to the (Vlasov) advection  term. The r.h.s. takes into account correlations (finite $N$ effects, graininess, discreteness effects) between stars that develop due to their interactions. These correlations correspond to ``collisions''.

Equation (\ref{gen2}) may be viewed as  a linear first order differential equation in time.  It can be symbolically written as
\begin{eqnarray}
\label{gen3} \frac{\partial g}{\partial t}+{\cal L}g={\cal S}\lbrack f \rbrack,
\end{eqnarray}
where  ${\cal L}={\cal L}_0+\langle{\cal L}\rangle$ is a mean field Liouvillian  and ${\cal S}[f]$ is a source term ${\cal S}$ expressible in terms of the one-body distribution.  This equation can be solved by the method of characteristics. Introducing the Green function
\begin{eqnarray}
\label{gen4}
G(t,t')={\rm exp}\left\lbrace -\int_{t'}^{t}{\cal L}(\tau)\, d\tau\right\rbrace,
\end{eqnarray}
constructed with the mean field Liouvillian ${\cal L}$, we obtain
\begin{eqnarray}
\label{gen5} g({\bf x},{\bf x}_1,t)=-\frac{1}{m^{2}}\int_0^t d\tau\, G(t,t-\tau)\nonumber\\
\times \left\lbrack
 \tilde{\bf F}(1\rightarrow 0)\cdot \frac{\partial}{\partial {\bf v}}+\tilde{\bf F}(0\rightarrow 1)\cdot \frac{\partial}{\partial {\bf v}_1}\right\rbrack \nonumber\\
 \times f({\bf x},t-\tau)f({\bf x}_1,t-\tau),
\end{eqnarray}
where we have assumed that no correlation is present initially so that $g({\bf x},{\bf x}_1,t=0)=0$ (if correlations are present initially, it can be shown that they are rapidly washed out). Substituting Eq. (\ref{gen5})  in Eq. (\ref{gen1}), we obtain
\begin{eqnarray}
\frac{\partial f}{\partial t}+{\bf v}\cdot {\partial f\over\partial {\bf r}}+\frac{N-1}{N}\langle {\bf F}\rangle\cdot {\partial f\over \partial {\bf v}}\nonumber\\
=\frac{\partial}{\partial
{v}^{\mu}}\int_0^t d\tau \int d{\bf r}_{1}d{\bf v}_1
{F}^{\mu}(1\rightarrow
0)G(t,t-\tau)\nonumber\\
\times  \left \lbrack {{\tilde F}}^{\nu}(1\rightarrow 0) {\partial\over\partial { v}^{\nu}}+{{\tilde F}}^{\nu}(0\rightarrow 1) {\partial\over\partial {v}_{1}^{\nu}}\right \rbrack \nonumber\\
{f}({\bf r},{\bf v},t-\tau)
\frac{f}{m}({\bf r}_1,{\bf
v}_1,t-\tau). \label{gen6}
\end{eqnarray}
In writing this equation, we have adopted a Lagrangian point of view. 
The coordinates ${\bf r}_i$ appearing after the Greenian must be viewed as
${\bf r}_i(t-\tau)={\bf r}_i(t)-\int_0^{\tau}ds\, {\bf v}_i(t-s)\, ds$
and  ${\bf v}_i(t-\tau)={\bf v}_i(t)-\int_0^{\tau}ds\, \langle {\bf
F}\rangle({\bf r}_i(t-s),t-s)\, ds$. Therefore, in order to evaluate the integral (\ref{gen6}), 
we must move the stars following the trajectories determined by the self-consistent mean field.

The  kinetic equation (\ref{gen6}) is valid at the order
$1/N$ so it describes the ``collisional'' evolution of the system
(ignoring collective effects) on a timescale of order $Nt_D$. Equation (\ref{gen6}) is a
non-Markovian integro-differential equation. It takes into account delocalizations in
space and time (i.e. spatial inhomogeneity and memory effects). Actually,
the Markovian approximation is justified
in the $N\rightarrow +\infty$ limit because the timescale $\sim Nt_D$
over which $f({\bf r},{\bf v},t)$ changes is long compared to the correlation
time $\tau_{corr}\sim t_D$ over which the integrand in Eq. (\ref{gen6}) has
significant support\footnote{It is sometimes argued that the Markovian approximation is not justified for stellar systems because the force auto-correlation decreases slowly as $1/t$ (Chandrasekhar 1944). However, this result is only true for an infinite homogeneous system (see Appendix \ref{sec_tcf}). For spatially inhomogeneous distributions, the correlation function decreases more rapidly and the Markovian approximation is justified (Severne and Haggerty 1976).}. Therefore, we can compute the correlation function (\ref{gen5}) by assuming that the distribution function is ``frozen'' at time $t$. This corresponds to the Bogoliubov ansatz in plasma physics.  If we replace $f({\bf r},{\bf v},t-\tau)$ and $f({\bf r}_1,{\bf v}_1,t-\tau)$ by $f({\bf r},{\bf v},t)$ and $f({\bf r}_1,{\bf v}_1,t)$ in Eq. (\ref{gen6}) and extend the time integral to
$+\infty$, we obtain
\begin{eqnarray}
\frac{\partial f}{\partial t}+{\bf v}\cdot {\partial f\over\partial {\bf r}}+\frac{N-1}{N}\langle {\bf F}\rangle \cdot {\partial f\over \partial {\bf v}}\nonumber\\
=\frac{\partial}{\partial
{v}^{\mu}}\int_0^{+\infty} d\tau \int d{\bf r}_{1}d{\bf v}_1
{F}^{\mu}(1\rightarrow
0)\nonumber\\
\times  G(t,t-\tau)\left \lbrack {{\tilde F}}^{\nu}(1\rightarrow 0) {\partial\over\partial { v}^{\nu}}+{{\tilde F}}^{\nu}(0\rightarrow 1) {\partial\over\partial {v}_{1}^{\nu}}\right \rbrack \nonumber\\
{f}({\bf r},{\bf v},t)\frac{f}{m}({\bf r}_1,{\bf
v}_1,t). \label{gen7}
\end{eqnarray}
Similarly, we can compute the trajectories of the stars by assuming that the mean field is independent on $\tau$ and equal to its value at time $t$ so that  ${\bf r}_i(t-\tau)={\bf r}_i(t)-\int_0^{\tau}ds\, {\bf v}_i(t-s)\, ds$
and  ${\bf v}_i(t-\tau)={\bf v}_i(t)-\int_0^{\tau}ds\, \langle {\bf
F}\rangle({\bf r}_i(t-s),t)\, ds$.

The structure of the kinetic equation (\ref{gen7}) has a clear physical meaning. The l.h.s. corresponds to the Vlasov advection term due to mean field effects. The r.h.s. can be viewed as a collision operator $C_N[f]$ taking finite $N$ effects into account.  For $N\rightarrow +\infty$, it vanishes and we recover the Vlasov equation. For finite $N$, it describes the cumulative effect of  binary collisions between stars. The collision operator is a sum of two terms: A diffusion term and a friction term. The coefficients of diffusion and friction are given by generalized Kubo formulae, i.e. they involve the time integral of the auto-correlation function of the fluctuating force. The kinetic equation (\ref{gen7}) bears some resemblance with the Fokker-Planck equation. However, it is more complicated since it is an {\it integro-differential} equation, not a differential equation (see Section \ref{sec_tp}).

Equation (\ref{gen7}) may be viewed as a generalized Landau
equation. Since the spatial inhomogeneity and the finite extension of
the system are properly taken into account, there is no divergence at
large scales. There is, however, a logarithmic divergence at small
scales which is due to the neglect of the interaction term ${\cal
L}_{12}$ in the Liouvillian (see Sec. \ref{sec_reg}). At large scales
(i.e. for large impact parameters), this term can be neglected and the
trajectories of the stars are essentially due to the mean field. Thus,
we can make the weak coupling approximation leading to the Landau
equation. This approximation describes {\it weak} collisions. However,
at small scales (i.e. for small impact parameters), we cannot ignore
the interaction term ${\cal L}_{12}$ in the Liouvillian anymore and we
have to solve the classical two-body problem. This is necessary to
correctly describe {\it strong} collisions for which the trajectory of
the particles deviates strongly from the mean field motion. When the
mean field Greenian $G$ (constructed with ${\cal L}_0+\langle{\cal
L}\rangle$) is replaced by the total Greenian $G'$ (constructed with
${\cal L}_0+{\cal L}_{12}+\langle{\cal L}\rangle$), taking into
account the interaction term, the generalized Landau equation
(\ref{gen7}) is replaced by a more complicated equation which can be
viewed as a generalized Boltzmann equation. This equation is free of
divergence (since it takes both spatial inhomogeneity and strong
collisions into account) but it is unnecessarily complicated because
it does not exploit the dominance of weak collisions over (rare) strong
collisions for the gravitational potential. Indeed, a star suffers a
large number of weak distant encounters and very few close
encounters. A better practical procedure is to use the generalized
Landau equation (\ref{gen7}) with a cut-off at small scales in order
to take into account our inability to describe strong collisions by this
approach.

The generalized Landau equation (\ref{gen7}) was derived by Kandrup
(1981) from the Liouville equation by using the projection operator
formalism. It can also be derived from the Liouville equation by
using the BBGKY hierarchy or from the Klimontovich equation by making a
quasilinear approximation (Chavanis 2008).

\subsection{The Vlasov-Landau equation}
\label{sec_vl}

Self-gravitating systems are intrinsically spatially
inhomogeneous. However, the collision operator at position ${\bf r}$
can be simplified by making a {\it local approximation} and performing
the integrals {\it as if} the system were spatially homogeneous with
the density $\rho=\rho({\bf r})$. This amounts to replacing $f({\bf
r}_{1},{\bf v}_{1},t)$ by $f({\bf r},{\bf v}_{1},t)$ and $\tilde{\bf
F}(i\rightarrow j)$ by ${\bf F}(i\rightarrow j)$ in
Eq. (\ref{gen7}). This local approximation is motivated by the work of
Chandrasekhar and von Neumann (1942) who showed that the distribution
of the gravitational force is a L\'evy law (called the Holtzmark
distribution) dominated by the contribution of the nearest
neighbor. With this local approximation, Eq. (\ref{gen7}) becomes
\begin{eqnarray}
\frac{\partial f}{\partial t}+{\bf v}\cdot {\partial f\over\partial {\bf r}}+\frac{N-1}{N}\langle {\bf F}\rangle \cdot {\partial f\over \partial {\bf v}}\nonumber\\
=\frac{\partial}{\partial
{v}^{\mu}}\int_0^{+\infty} d\tau \int d{\bf r}_{1}d{\bf v}_1
{F}^{\mu}(1\rightarrow
0)G_0(t,t-\tau)\nonumber\\
\times  {{F}}^{\nu}(1\rightarrow 0)\left (  {\partial\over\partial { v}^{\nu}}- {\partial\over\partial {v}_{1}^{\nu}}\right ){f}({\bf r},{\bf v},t)\frac{f}{m}({\bf r},{\bf
v}_1,t), \qquad\label{vl0}
\end{eqnarray}
where we have used ${\bf F}(0\rightarrow 1)=-{\bf F}(1\rightarrow 0)$. The Greenian ${G}_0$ corresponds to the free motion of the particles associated with the Liouvillian ${\cal L}_0$. Using Eqs. (\ref{sw5a}) and (\ref{sw5b}), the foregoing equation can be rewritten as
\begin{eqnarray}
\frac{\partial f}{\partial t}+{\bf v}\cdot {\partial f\over\partial {\bf r}}+\frac{N-1}{N}\langle {\bf F}\rangle \cdot {\partial f\over \partial {\bf v}}\nonumber\\
=\frac{\partial}{\partial
{v}^{\mu}}\int_0^{+\infty} d\tau \int d{\bf r}_{1}d{\bf v}_1
{F}^{\mu}(1\rightarrow
0,t){{F}}^{\nu}(1\rightarrow 0,t-\tau)\nonumber\\
\times  \left (  {\partial\over\partial { v}^{\nu}}- {\partial\over\partial {v}_{1}^{\nu}}\right ){f}({\bf r},{\bf v},t)\frac{f}{m}({\bf r},{\bf
v}_1,t), \qquad\label{vl1}
\end{eqnarray}
where ${\bf F}(1\rightarrow 0,t-\tau)$ is expressed in terms of the Lagrangian coordinates.  The integrals over $\tau$ and ${\bf r}_{1}$ can be calculated explicitly (see Appendix \ref{sec_sw}). We then find that the evolution of the distribution function is governed by the
Vlasov-Landau equation
\begin{eqnarray}
\frac{\partial f}{\partial t}+{\bf v}\cdot
{\partial f\over\partial {\bf r}}+\frac{N-1}{N}\langle {\bf F}\rangle\cdot {\partial
f\over\partial {\bf v}}=\pi (2\pi)^3 m\nonumber\\
\times\frac{\partial}{\partial {v}^{\mu}} \int
k^{\mu} k^{\nu} \delta ({\bf k}\cdot {\bf w})\hat{u}^2({k})
 \left (f_1\frac{\partial
f}{\partial v^{\nu}}-f\frac{\partial f_1}{\partial
v_1^{\nu}}\right )\, d{\bf v}_1 d{\bf k}, \label{vl2}
\end{eqnarray}
where we have noted ${\bf w}={\bf v}-{\bf v}_1$, $f=f({\bf r},{\bf
v},t)$, $f_1=f({\bf r},{\bf v}_1,t)$, and where $(2\pi)^3\hat{u}(k)=-4\pi G/k^2$ represents the Fourier transform of the gravitational potential. Under this form, we see that the collisional evolution of a stellar system is due to a condition of resonance ${\bf k}\cdot {\bf v}={\bf k}\cdot {\bf v}'$ (with ${\bf v}\neq {\bf v}'$) encapsulated in the $\delta$-function. This $\delta$-function expresses the conservation of energy.

The Vlasov-Landau equation can also be written as (see Appendix \ref{sec_sw}):
\begin{eqnarray}
\label{vl3} \frac{\partial f}{\partial t}+{\bf v}\cdot
{\partial f\over\partial {\bf r}}&+&\frac{N-1}{N}\langle {\bf F}\rangle\cdot {\partial
f\over\partial {\bf v}}\nonumber\\
&=&{\partial\over\partial
v^{\mu}}\int  K^{\mu\nu}\biggl (f_1{\partial
f\over\partial v^{\nu}}-f {\partial
f_1\over\partial v^{\nu}_1}\biggr )\, d{\bf v}_1,
\end{eqnarray}
\begin{equation}
\label{vl4}
K^{\mu\nu}=A\frac{w^2\delta^{\mu\nu}-{w^{\mu}w^{\nu}}}{w^3},\qquad A=2\pi m G^{2}\ln\Lambda,
\end{equation}
where
\begin{equation}
\label{vl5}
\ln
\Lambda=\int_{k_{min}}^{k_{max}}\, dk/k,
\end{equation}
is the Coulomb factor that has to be regularized with appropriate
cut-offs (see Section \ref{sec_reg}). The r.h.s. of Eq. (\ref{vl3}) is
the original form of the collision operator given by Landau (1936) for
the Coulombian interaction\footnote{The case of plasmas is recovered
by making the substitution $Gm^{2}\leftrightarrow -e^{2}$ leading to
$K=(2\pi e^4/m^3)\ln\Lambda$.}. It applies to weakly coupled
plasmas. We note that the potential of interaction only appears in the
constant $A$ which merely determines the relaxation time. The
structure of the Landau equation is independent on the potential. The
Landau equation was originally derived from the Boltzmann equation in
the limit of weak deflections $|\Delta {\bf v}|\ll 1$ (Landau
1936)\footnote{We note that Eq. (\ref{vl0}) with $G_0$ replaced by the
total Greenian ${G}'$ taking into account the interaction term is
equivalent to the Boltzmann equation. Indeed, for spatially
homogeneous systems, the Boltzmann equation can be derived from the
BBGKY hierarchy (\ref{trunc5}) and (\ref{trunc6}) by keeping the
Liouvillian ${\cal L}={\cal L}_0+{\cal L}_{12}$ describing the
two-body problem exactly and the source ${\cal S}$ (Balescu
2000). Therefore, the procedure used by Landau which amounts to
expanding the Boltzmann equation in the limit of weak deflections is
equivalent to the one presented here that starts from the BBGKY
hierarchy and neglects ${\cal L}_{12}$ in the Liouvillian.}.  In the
case of plasmas, the system is spatially homogeneous and the advection
term is absent in Eq. (\ref{vl3}). In the case of stellar systems,
when we make the local approximation, the spatial inhomogeneity of the
system is only retained in the advection term of Eq. (\ref{vl3}). This
is why this kinetic equation is referred to as the Vlasov-Landau
equation. This is the fundamental kinetic equation of stellar systems.

\subsection{Heuristic regularization of the divergence}
\label{sec_reg}

To obtain the Vlasov-Landau  equation (\ref{vl3}), we have made a local approximation. This amounts to calculating the collision operator at each point as if the system were spatially homogeneous. As a result of this homogeneity assumption, a logarithmic  divergence appears at large scales in the Coulombian factor (\ref{vl5}). In  plasma physics, this divergence is cured  by the Debye shielding. A charge is surrounded by a polarization cloud of opposite charges which reduces the range of the interaction. When collective effects are properly taken into account, as in the Lenard-Balescu equation, no divergence appears as large scales and the Debye length arises naturally. Heuristically, we can use the Landau equation and introduce an upper cut-off at the Debye length $\lambda_D=(k_B T/ne^2)^{1/2}$. For self-gravitating systems, there is no shielding and the divergence is cured by the finite extent of the system. The interaction between two stars is only limited by the size of the system. When spatial inhomogeneity is taken into account, as  in the generalized Landau equation (\ref{gen7}), no divergence occurs at large scales. Heuristically, we can use the Vlasov-Landau equation (\ref{vl3}) and introduce an upper cut-off at the Jeans length $\lambda_J\sim (k_BT/Gm^2 n)^{1/2}$  which gives an estimate of the system size $R$.

The Coulombian factor (\ref{vl5})  also diverges at small scales. As explained previously, this is due to the neglect of strong collisions that produce important deflections.  Indeed, for collisions with low impact parameter, the mean field approximation is clearly irrelevant and it is necessary to solve the two-body problem exactly. Heuristically, we can use the Landau equation and introduce a lower cut-off at the gravitational Landau length  $\lambda_L\sim Gm/v_m^2\sim Gm^2/(k_B T)$ (the gravitational analogue of the Landau length $\lambda_L\sim e^2/m v_m^2\sim e^2/k_B T$ in plasma physics) which corresponds to the impact parameter leading to a deflection at $90^{o}$.

Introducing a large-scale cut-off at the Jeans length $\lambda_J$ and a small-scale cut-off at the Landau length $\lambda_L$, and noting that  $\lambda_L\sim 1/(n\lambda_J^2)$, we find that the Coulombian factor scales as $\ln\Lambda\sim \ln
({\lambda_J}/{\lambda_L})\sim \ln (n\lambda_J^3)\sim \ln N$ where $N\sim n\lambda_J^3\sim nR^3$ is the number of stars in the
cluster\footnote{Similarly, in plasma physics, introducing a large-scale cut-off at the Debye length $\lambda_D$ and a small-scale cut-off at the Landau length $\lambda_L$, and noting that  $\lambda_L\sim 1/(n\lambda_D^2)$, we find that the Coulombian factor scales as $\ln\Lambda\sim \ln
({\lambda_D}/{\lambda_L})\sim \ln (n\lambda_D^3)$ where $\Lambda\sim n\lambda_D^3$ is the number of electrons in the Debye sphere.}.

\subsection{Properties of the Vlasov-Landau equation}
\label{sec_prop}

The Vlasov-Landau equation conserves the mass $M=\int f\, d{\bf
r}d{\bf v}$ and the energy $E=\int f\frac{v^2}{2}\, d{\bf r}d{\bf
v}+\frac{1}{2}\int\rho\Phi\, d{\bf r}$. It also monotonically
increases the Boltzmann entropy $S=-\int ({f}/{m})\ln
({f}/{m})\, d{\bf r}d{\bf v}$ in the sense that $\dot S\ge 0$ ($H$-theorem). Due to the local approximation, the proof of these properties is the same\footnote{We recall that the Vlasov advection term conserves mass, energy, and entropy.} as for the spatially homogeneous Landau equation (Balescu 2000). Because of these properties, we might expect that a stellar system will relax towards the Boltzmann distribution which maximizes the entropy at fixed mass and energy. However, we know that there is no maximum entropy state for an unbounded self-gravitating system (the Boltzmann distribution has infinite mass). Therefore,  the  Vlasov-Landau equation does not relax towards a steady state and the entropy does not reach a stationary value. Actually, the entropy increases permanently as the system evaporates. But since evaporation is a slow process, the system may achieve a quasistationary state that is {\it close} to the Boltzmann distribution. A typical quasistationary distribution is the Michie-King model
\begin{eqnarray}
f=A e^{-\beta m j^2/(2r_a^2)}(e^{-\beta m\epsilon}-e^{-\beta m\epsilon_m}), 
\label{mk}
\end{eqnarray} 
where $\epsilon=v^2/2+\Phi({\bf r})$ is the energy and ${\bf j}={\bf r}\times {\bf v}$ the angular momentum. This distribution takes into account the escape of high energy stars and the anisotropy of the velocity distribution. It can be derived, under some approximations, from the Vlasov-Landau equation by using the condition that $f=0$ if the energy of the star is larger than the escape energy $\epsilon_m$ (Michie 1963, King 1965). The Michie-King distribution reduces to the isothermal distribution $f\propto e^{-\beta m\epsilon}$ for low energies.  In this sense, we can define a ``relaxation'' time for a stellar system. From the Vlasov-Landau equation (\ref{vl3}), we find that the relaxation time scales as
\begin{eqnarray}
t_R\sim \frac{v_m^3}{nm^2G^2\ln N},
\label{prop1}
\end{eqnarray}
where $v_m$ is the root mean square (r.m.s.) velocity.
Introducing the dynamical time $t_D\sim \lambda_J/v_m\sim R/v_m$, we obtain the scaling
\begin{eqnarray}
t_R\sim \frac{N}{\ln N}t_D.
\label{prop2}
\end{eqnarray}
The fact that the ratio between the relaxation time and the dynamical time depends only on the number of stars and scales as $N/\ln N$ was first noted by Chandrasekhar (1942).

A simple estimate of the evaporation time gives $t_{evap}\simeq 136 t_R$  (Ambartsumian 1938, Spitzer 1940). More precise values have been obtained by studying the evaporation process in an artificially uniform cluster (Chandrasekhar 1943b, Spitzer and H\"arm 1958, Michie 1963, King 1965, Lemou and Chavanis 2010) or in a more realistic inhomogeneous cluster (H\'enon 1961). Since $t_{evap}\gg t_R$, we can consider that the system  relaxes towards a steady distribution of the form (\ref{mk}) on a timescale $t_R$ and that this distribution slowly evolves on a longer timescale as the stars escape\footnote{The distribution (\ref{mk}) is not steady in the sense that the coefficients $A$, $\beta$, and $r_a$ slowly vary in time as the stars escape and the system loses mass and energy. However, the distribution keeps the same form during the evaporation process. H\'enon (1961) found a self-similar solution of the orbit-averaged-Fokker-Planck equation. He showed that the core contracts as the cluster evaporates and that the central density is infinite (the structure of the core resembles the singular isothermal sphere with a density $\rho\propto r^{-2}$). He argued that the concentration of energy, without concentration of mass, at the center of the system is due to the formation of tight binary stars. The invariant profile found by H\'enon does not exactly coincide with the Michie-King distribution (introduced later), but is reasonably close.}. The characteristic time in which the system's stars evaporate is $t_{evap}$. The evaporation is one reason for the evolution of stellar systems. However, as demonstrated by Antonov (1962) and Lynden-Bell and Wood (1968), stellar systems may evolve more rapidly as a result of the gravothermal catastrophe. In that case, the Michie-King distribution changes significatively due to core collapse. This evolution has been described by Cohn (1980), and it leads ultimately to the formation of a binary star surrounded by a hot halo. Such a configuration can have arbitrarily large entropy. Cohn (1980) finds that the entropy increases permanently during core collapse, confirming that the Vlasov-Landau equation has no equilibrium state.

Even if the system were confined within a small box so as to prevent both the evaporation and the gravothermal catastrophe, there would be no statistical equilibrium state in a strict sense because there is no {\it global} entropy maximum (Antonov 1962). A configuration in which some subset of the particles are tightly bound together (e.g. a binary star), and in which the rest of the particles shares the energy thereby released, may have an arbitrary large entropy. However, such configurations, which require strong correlations, are generally  reached very slowly (on a timescale much larger than $(N/\ln N)t_D$) due to encounters involving many particles. To describe these configurations, one would have to take high order correlations into account in the kinetic theory. These configurations may be relevant in systems with a small number of stars (Chabanol et al. 2000) but when $N$ is large the picture is different. On a timescale of the order of $(N/\ln N)t_D$ the one-body distribution function is expected to reach the Boltzmann distribution which is a {\it local} entropy maximum. This state is ``metastable'' but its lifetime is expected to be very large, scaling as $e^N$,  so that it is stable in practice (Chavanis 2006). In this sense, there exist ``true'' statistical equilibrium states for self-gravitating systems confined within a small box. However, we may argue that  this situation is highly artificial.

\subsection{Dynamical evolution of stellar systems: a short review}
\label{sec_shr}

Using the kinetic theory, we can identify different phases in the dynamical evolution of stellar systems.

A self-gravitating system initially out-of-mechanical equilibrium
undergoes a process of {\it violent collisionless relaxation} towards
a virialized state. In this regime, the
dynamical evolution of the cluster is described by the Vlasov-Poisson
system. The phenomenology of violent relaxation has been described
by H\'enon (1964), King (1966), and Lynden-Bell (1967). Numerical simulations that start from cold and clumpy
initial conditions generate a quasi stationary state (QSS) that fits
the de Vaucouleurs $R^{1/4}$ law for the surface brightness of elliptical galaxies 
quite well (van Albada 1982). The
inner core is almost isothermal (as predicted by Lynden-Bell 1967) while 
the velocity distribution in the envelope
is radially anisotropic and the density profile decreases as
$r^{-4}$. One success of Lynden-Bell's statistical theory
of violent relaxation is to explain the isothermal core of elliptical galaxies without
recourse to ``collisions''. By contrast, the structure of the halo
cannot be explained by Lynden-Bell's theory as it results from an
{\it incomplete relaxation}. Models of incompletely relaxed stellar systems have been elaborated by Bertin and Stiavelli (1984), Stiavelli and Bertin (1987), and Hjorth and Madsen (1991). These theoretical models nicely reproduce the results of observations or numerical simulations 
(Londrillo et al. 1991, Trenti et al. 2005). In the simulations, the  initial condition needs to be sufficiently clumpy and cold to generate enough mixing required for a successful application of the statistical theory of violent relaxation. Numerical simulations
starting from homogeneous spheres (see, e.g., Roy and Perez 2004, Levin et al. 2008, Joyce et al. 2009)
show little angular momentum mixing and lead to different results. In
particular, they display a larger amount of mass loss (evaporation)
than simulations starting from clumpy initial conditions. Clumps thus
help the system to reach a ``universal'' final state from a variety of
initial conditions, which can explain the similarity of the density
profiles observed in elliptical galaxies.

On longer timescales, encounters between stars must be taken into account and the dynamical evolution of the cluster is governed by the Vlasov-Landau-Poisson system.
The first stage of the collisional evolution is driven by
{\it evaporation}. Due to a series of weak encounters, the energy of a
star can gradually increase until it reaches the local escape energy;
in that case, the star leaves the system.
 Numerical simulations (Spitzer 1987, Binney and Tremaine 2008) show that
during this regime the system reaches a quasi-stationary state that
slowly evolves in amplitude due to evaporation as the system loses
mass and energy. This quasi stationary distribution function is
close to the Michie-King model (\ref{mk}). The system has a core-halo
structure. The core is isothermal while the stars in the outer halo
move in predominantly radial orbits. Therefore, the distribution
function in the halo is anisotropic.  The density follows the
isothermal law $\rho\sim r^{-2}$ in the central region (with a core of
almost uniform density) and decreases as $\rho\sim r^{-7/2}$ in the
halo (for an isolated cluster with $\epsilon_m=0$). Due to evaporation, the halo expands while the core shrinks as required by energy conservation. During the evaporation process, the
central density increases permanently. At some point of the
evolution, the system undergoes an
instability related to the Antonov (1962) instability\footnote{The collisional evolution of the system can be measured
precisely in terms of the scale escape energy
$x_0=3\Phi(0)/v_m^2(0)$. The onset of instability corresponds to $x_0\simeq 9.3$ which is the value at which the King model becomes thermodynamically unstable (Cohn 1980).}   and the
gravothermal catastrophe sets in (Lynden-Bell and Wood 1968). This instability is due to the negative specific heat of
the inner system that evolves by losing energy and thereby growing
hotter. The energy lost is transferred outward by stellar encounters. Hence the temperature always decreases outward, and the center continually loses energy, shrinks, and heats up. This leads to {\it core
collapse}. Mathematically speaking, core collapse would
generate a finite time singularity. When the evolution is modeled by the
orbit-averaged-Fokker-Planck equation, Cohn (1980) finds that the
collapse is self-similar, that the central density becomes infinite in
a finite time, and that the density behaves as $\rho\sim r^{-2.23}$. The invariant profile found by Cohn differs from the Michie-King distribution (for which $\rho\sim r^{-2}$)
beyond a radius of about $10 r_{core}$. Larson (1970) and Lynden-Bell and  Eggleton (1980) find similar results by modeling the evolution of the system by fluid equations.
Alternatively, when the evolution is modeled by the original Vlasov-Landau equation,
Lancellotti and Kiessling (2001) argue that the density should behave as $\rho\propto r^{-3}$
in the final stage of the collapse. In all cases, the authors find a singular density profile at the collapse time that is integrable (or diverges logarithmically) at $r=0$. This means that the core contains very little mass. In reality, if we come back to
the $N$-body system, there is no singularity and  core collapse is arrested by the formation of
binary stars due to three-body collisions. These binaries can release sufficient energy to stop the
collapse  and even drive a re-expansion of the cluster in a
post-collapse regime (Inagaki and Lynden-Bell 1983). Then, in principle, a series of gravothermal
oscillations should follow (Bettwieser and Sugimoto 1984).

At the present epoch, small groups of stars such as globular clusters ($N\sim
10^{5}$, $t_D\sim 10^5\, {\rm yr}$, age $\sim 10^{10}\, {\rm yr}$, $t_R\sim 10^{10}\, {\rm yr}$) are in the collisional regime. They are either
in  quasistationary states described by the Michie-King model or experiencing core collapse. By contrast, large clusters of stars like elliptical
galaxies ($N\sim 10^{11}$, $t_D\sim 10^8\, {\rm yr}$, age $\sim 10^{10}\, {\rm yr}$, $t_R\sim 10^{19}\, {\rm yr}$) are still in the collisionless regime and their apparent
organization is a result of an incomplete violent relaxation.

\section{Test star in a thermal bath}
\label{sec_tp}

\subsection{The Fokker-Planck equation}
\label{sec_fp}

We now consider the relaxation of a ``test'' star (tagged particle)
evolving in a steady distribution of ``field'' stars\footnote{In plasma physics, this steady distribution is the Boltzmann distribution of statistical equilibrium which is the steady state of the Landau equation. In the case of stellar systems, there is an intrinsic difficulty since no statistical equilibrium state exists in a strict sense: The Boltzmann distribution has infinite mass, and the Vlasov-Landau equation has no steady state because of the escape of high energy stars. However, we have seen that the system can reach a quasi steady distribution (e.g. a Michie-King distribution) and that this distribution changes on an evaporation timescale that is long with respect to the collisional relaxation time. Therefore, we can consider that this distribution is steady on the collisional timescale over which the Fokker-Planck approach applies.}. Due to the encounters with the
field stars, the test star has
a stochastic motion.  We call
$P({\bf r},{\bf v},t)$ the probability density of finding the test
star at position ${\bf r}$ with velocity ${\bf v}$ at time $t$. The
evolution of $P({\bf r},{\bf v},t)$ can be obtained from the generalized Landau
equation (\ref{gen7}) by considering that the distribution function of
the field stars is {\it fixed}. Therefore, we replace
$f({\bf r},{\bf v},t)$ by $P({\bf r},{\bf v},t)$ and
$f({\bf r}_1,{\bf v}_{1},t)$ by $f({\bf r}_1,{\bf v}_1)$
where $f({\bf r}_1,{\bf v}_1)$ is the steady distribution of the field stars\footnote{We can understand this procedure as follows. Equations (\ref{gen7}) and (\ref{fpnew2}) govern the evolution of the distribution function of a test star (described by the coordinates ${\bf r}$ and ${\bf v}$) interacting with field stars (described by the running coordinates ${\bf r}_1$ and ${\bf v}_1$). In Eq. (\ref{gen7}), all the stars are equivalent so the distribution of the field stars $f({\bf r}_1,{\bf v}_1,t)$ changes with time exactly like the distribution of the test star $f({\bf r},{\bf v},t)$. In Eq. (\ref{fpnew2}), the test star and the field stars are not equivalent since the field stars form a ``bath''. The field stars have a steady (given) distribution $f({\bf r}_1,{\bf v}_1)$ while the distribution of the test star $f({\bf r},{\bf v},t)=N m P({\bf r},{\bf v},t)$ changes with time.}.  This procedure transforms the integro-differential equation
(\ref{gen7}) into the differential equation
\begin{eqnarray}
\frac{\partial P}{\partial t}+{\bf v}\cdot {\partial P\over\partial {\bf r}}+\frac{N-1}{N}\langle {\bf F}\rangle \cdot {\partial P\over \partial {\bf v}}\nonumber\\
=\frac{\partial}{\partial
{v}^{\mu}}\int_0^{+\infty} d\tau \int d{\bf r}_{1}d{\bf v}_1
{F}^{\mu}(1\rightarrow
0)\nonumber\\
\times  G(t,t-\tau)\left \lbrack {{\tilde F}}^{\nu}(1\rightarrow 0) {\partial\over\partial { v}^{\nu}}+{{\tilde F}}^{\nu}(0\rightarrow 1) {\partial\over\partial {v}_{1}^{\nu}}\right \rbrack \nonumber\\
{P}({\bf r},{\bf v},t)\frac{f}{m}({\bf r}_1,{\bf
v}_1). \label{fpnew2}
\end{eqnarray}
where $\langle {\bf F}\rangle({\bf r})=-\nabla\Phi({\bf r})$  is the {\it static} mean force created by the field stars with density $\rho({\bf r}_1)$. This equation does not present any divergence at large scales.

If we make a local approximation and use the Vlasov-Landau equation (\ref{vl1}), we obtain 
\begin{eqnarray}
\frac{\partial P}{\partial t}+{\bf v}\cdot {\partial P\over\partial {\bf r}}+\frac{N-1}{N}\langle {\bf F}\rangle \cdot {\partial P\over \partial {\bf v}}\nonumber\\
=\frac{\partial}{\partial
{v}^{\mu}}\int_0^{+\infty} d\tau \int d{\bf r}_{1}d{\bf v}_1
{F}^{\mu}(1\rightarrow
0,t){{F}}^{\nu}(1\rightarrow 0,t-\tau)\nonumber\\
\times  \left (  {\partial\over\partial { v}^{\nu}}- {\partial\over\partial {v}_{1}^{\nu}}\right ){P}({\bf r},{\bf v},t)\frac{f}{m}({\bf r},{\bf
v}_1), \qquad\label{fpnew1}
\end{eqnarray}
Denoting the advection operator by $d/dt$, Eq. (\ref{fpnew1}) can be written in the form of a Fokker-Planck
equation
\begin{equation}
\label{fp2}
\frac{dP}{dt}={\partial\over\partial v^{\mu}}\biggl (D^{\mu\nu}{\partial P\over\partial v^{\nu}}-P F_{pol}^{\mu}\biggr ),
\end{equation}
involving a diffusion tensor
\begin{eqnarray}
\label{fpnew3} D^{\mu\nu}=\frac{1}{m}\int_0^{+\infty} d\tau \int d{\bf r}_{1}d{\bf v}_1
{F}^{\mu}(1\rightarrow
0,t)\nonumber\\
\times{{F}}^{\nu}(1\rightarrow 0,t-\tau)f({\bf r},{\bf
v}_1), 
\end{eqnarray}
and a friction force
\begin{eqnarray}
F_{pol}^{\mu}=\frac{1}{m}\int_0^{+\infty} d\tau \int d{\bf r}_{1}d{\bf v}_1
{F}^{\mu}(1\rightarrow 0,t)\nonumber\\
\times
{{F}}^{\nu}(1\rightarrow 0,t-\tau) {\partial f\over\partial {v}_{1}^{\nu}}({\bf r},{\bf
v}_1).
\label{fpnew4}
\end{eqnarray}

If we directly start from Eq. (\ref{vl2}), which amounts to performing the integrals over $\tau$ and ${\bf r}_1$ in the previous expressions, we obtain the Fokker-Planck equation
\begin{eqnarray}
\label{fp1} \frac{\partial P}{\partial t}+{\bf v}\cdot
{\partial P\over\partial {\bf r}}+\frac{N-1}{N}\langle {\bf F}\rangle\cdot {\partial
P\over\partial {\bf v}}
=\pi (2\pi)^{3}m\nonumber\\
\times{\partial\over\partial v^{\mu}}\int k^{\mu}k^{\nu}\delta ({\bf k}\cdot {\bf w})\hat{u}({k})^{2}\biggl (f_1 {\partial P \over\partial v^{\nu}}-P{\partial f_1\over\partial {v}_{1}^{\nu}}\biggr )\, d{\bf v}_{1}d{\bf k},
\end{eqnarray}
with the diffusion and friction coefficients
\begin{equation}
\label{fp3} D^{\mu\nu}=\pi (2\pi)^{3}m\int k^{\mu}k^{\nu}\delta ({\bf k}\cdot {\bf w})\hat{u}({k})^{2} f_1 \, d{\bf v}_{1}d{\bf k},
\end{equation}
\begin{eqnarray}
F_{pol}^{\mu}=\pi (2\pi)^{3}m\int k^{\mu}k^{\nu}\delta ({\bf k}\cdot {\bf w}) \hat{u}({k})^{2}{\partial f_1\over\partial {v}_{1}^{\nu}} \, d{\bf v}_{1}d{\bf k}.
\label{fp4}
\end{eqnarray}

The diffusion tensor $D^{\mu\nu}$ results from  the fluctuations
of the gravitational force due to the granularities in the distribution of the field stars. It can be derived directly from the formula (see Appendix \ref{sec_tcf}):
\begin{eqnarray}
\label{fpnew3b} D^{\mu\nu}=\int_0^{+\infty}  \langle {F}^{\mu}(t){F}^{\nu}(t-\tau)\rangle\, d\tau,
\end{eqnarray}
deduced from Eq. (\ref{fp6}-a).  The friction force ${\bf
F}_{pol}$ results from the retroaction of the field stars to the perturbation caused by the test star like in a polarization process. It can be derived
from a linear response theory (Marochnik 1968, Kalnajs 1971b, Kandrup
1983, Chavanis 2008). It will be called the ``friction by
polarization'' to distinguish it from the total friction (see below).
Eqs. (\ref{fpnew1})-(\ref{fp4}) have been derived within the
local approximation. More general formulae, valid for fully
inhomogeneous stellar systems, are given in Kandrup (1983) and
Chavanis (2008). The friction force has also been calculated by
Tremaine \& Weinberg (1984), Bekenstein \& Maoz (1992), Maoz (1993),
Nelson \& Tremaine (1999) using different approaches.

Since the diffusion tensor depends on the
velocity ${\bf v}$ of the test star, it is useful to rewrite
Eq. (\ref{fp2}) in a form that is fully consistent with the general
Fokker-Planck equation
\begin{equation}
\label{fp5} {dP\over d t}={\partial^{2}\over\partial v^{\mu}\partial v^{\nu}}(D^{\mu\nu}P)-{\partial\over\partial v^{\mu}}(P F_{friction}^{\mu}),
\end{equation}
with
\begin{eqnarray}
\label{fp6}
D^{\mu\nu}={\langle \Delta v^{\mu} \Delta v^{\nu}\rangle\over 2\Delta t}, \quad F_{friction}^{\mu}={\langle \Delta v^{\mu}\rangle\over \Delta t}.
\end{eqnarray}
By identification, we find that
\begin{eqnarray}
\label{fp7}
F_{friction}^{\mu}=F_{pol}^{\mu}+{\partial D^{\mu\nu}\over\partial v^{\nu}}.
\end{eqnarray}
Therefore, when the diffusion coefficient depends on the velocity, the
total friction is different from the friction by polarization.  Substituting
Eqs. (\ref{fp3}) and (\ref{fp4}) in Eq. (\ref{fp7}), and using an
integration by parts, we find that the diffusion and friction
coefficients can be written as
\begin{eqnarray}
\label{fp8} {\langle \Delta v^{\mu} \Delta v^{\nu}\rangle\over 2\Delta t}=\pi (2\pi)^{3}m\int k^{\mu}k^{\nu}\delta ({\bf k}\cdot {\bf w})\hat{u}({k})^{2} f_1 \, d{\bf v}_{1}d{\bf k},\nonumber\\
\end{eqnarray}
\begin{eqnarray}
{\langle \Delta v^{\mu}\rangle\over \Delta t}=\pi (2\pi)^{3}m\int k^{\mu}k^{\nu} f_1 \left (\frac{\partial}{\partial {v}^{\nu}}-\frac{\partial}{\partial v_1^\nu}\right ) \nonumber\\
\times\delta ({\bf k}\cdot {\bf w}) \hat{u}({k})^{2}\, d{\bf v}_{1}d{\bf k}.
\label{fp9}
\end{eqnarray}
These expressions can be obtained directly from the equations of motion by expanding the trajectories of the stars in powers of $1/N$ in the limit $N\rightarrow +\infty$ (Chavanis 2008). We recall that Eqs. (\ref{fp8}) and (\ref{fp9}) display a logarithmic divergence a small and large scales that must be regularized by introducing proper cut-offs as explained in Section \ref{sec_reg}. In astrophysics, the diffusion and friction coefficients of a star were first calculated by Chandrasekhar (1943a) from a two-body encounters theory (see also Cohen et al. 1950, Gasiorowicz et al. 1956, and Rosenbluth et al. 1957). The expressions obtained by these authors are different from those given above, but they are equivalent (see Section \ref{sec_rosen}). In plasma physics, the diffusion and friction coefficients of a charge were first calculated by Hubbard (1961a) who took collective effects into account, thereby eliminating the divergence at large scales. When collective effects are neglected, his expressions reduce to Eqs. (\ref{fp8}) and (\ref{fp9}). On the other hand, strong collisions have been taken into account by Chandrasekhar (1943a) in astrophysics and by Hubbard (1961b) in plasma physics. In that case, there is no divergence at small impact parameters in the diffusion and friction coefficients, and the Landau length appears naturally.

The two forms (\ref{fp2}) and (\ref{fp5}) of the Fokker-Planck equation have their own
interest. The expression (\ref{fp5}) where the diffusion coefficient
is placed {after} the two derivatives $\partial^{2}(DP)$ involves the
total friction force ${\bf F}_{friction}$ and the expression (\ref{fp2}) where the
diffusion coefficient is placed {between} the derivatives $\partial
D\partial P$ isolates the part of the friction ${\bf F}_{pol}$ due to the polarization.
Astrophysicists are used to the form (\ref{fp5}). However, it is the form (\ref{fp2}) that stems from the Landau equation (\ref{vl1}). We shall come back to this observation in Section \ref{sec_rosen}.

From Eqs. (\ref{fp3}) and (\ref{fp4}), we easily obtain
\begin{eqnarray}
\label{fp10}
\frac{\partial D^{\mu\nu}}{\partial v^\nu}=F_{pol}^\mu.
\end{eqnarray}
Combining Eq. (\ref{fp7}) with Eq. (\ref{fp10}), we get
\begin{equation}
\label{fp11}
{\bf F}_{friction}=2{\bf F}_{pol}.
\end{equation}
Therefore, the friction force ${\bf F}_{friction}$ is equal to twice the friction by polarization ${\bf F}_{pol}$ (for a test star with mass $m$ interacting with field stars with mass $m_f$, this factor two is replaced by $(m+m_f)/m$; see Appendix \ref{sec_ms}). This explains the difference of factor $2$ in the calculations of Chandrasekhar (1943a) who determined ${\bf F}_{friction}$ and in the calculations of Kalnajs (1971b) and Kandrup (1983) who determined ${\bf F}_{pol}$.

\subsection{The Einstein relation}
\label{sec_einstein}

In the central region of the system,  the distribution of the field stars is close to the Maxwell-Boltzmann distribution
\begin{equation}
\label{e1} f({\bf r}_1,{\bf v}_1)=\biggl ({\beta m\over 2\pi}\biggr )^{3/2}\rho({\bf r}_1) \ e^{-\beta m {v_1^{2}\over 2}},
\end{equation}
where $\beta=1/T$ is the inverse temperature and $\rho({\bf r}_1)\propto e^{-\beta m\Phi({\bf r}_1)}$ the density given by the Boltzmann law. Therefore, the field stars form a {\it thermal bath}. Substituting the identity
\begin{equation}
\label{e2}
\frac{\partial f_1}{\partial {\bf v}_1}=-\beta m f_1 {\bf v}_1,
\end{equation}
in Eq. (\ref{fp4}), using the $\delta$-function to replace ${\bf k}\cdot {\bf v}_1$ by ${\bf k}\cdot {\bf v}$, and comparing the resulting expression with Eq. (\ref{fp3}), we find that 
\begin{equation}
\label{e5} F^{\mu}_{pol}=-\beta m D^{\mu\nu}v^{\nu}.
\end{equation}
The friction coefficient is given by an Einstein relation expressing the fluctuation-dissipation theorem\footnote{The collisional evolution of a star, under the effect of two-body encounters, can be understood as an interplay between two competing effects. The fluctuations of the gravitational field induce a diffusion in velocity space which tends to increase the speed of the star. This effect is counterbalanced by a friction (dissipation) which results in a systematic deceleration along the direction of motion. As emphasized by Chandrasekhar (1943a, 1949) in his Brownian theory of stellar motion, the Einstein relation (\ref{e5}) guarantees that the Maxwell distribution (\ref{e1}) is a steady state of the Fokker-Planck equation (\ref{fp2}).}. We emphasize  that the Einstein relation is valid for the friction force by polarization ${\bf F}_{pol}$, not for the total friction ${\bf F}_{friction}$ (we do not have this subtlety in standard Brownian theory where the diffusion coefficient is constant). Using Eq. (\ref{fpnew3b}), we can rewrite Eq. (\ref{e5}) in the form
\begin{equation}
\label{enew5} F^{\mu}_{pol}=-\beta m v^{\nu}\int_0^{+\infty} d\tau \langle {F}^{\mu}(t){F}^{\nu}(t-\tau)\rangle.
\end{equation}
This relation is usually called  the Kubo formula. More general expressions of the Kubo formula valid for fully inhomogeneous stellar systems are given in Kandrup (1983) and Chavanis (2008). Using the Einstein relation, the Fokker-Planck equation
(\ref{fp2}) takes the form
\begin{eqnarray}
\label{e6}
\frac{dP}{dt}={\partial\over\partial v^{\mu}}\left\lbrack D^{\mu\nu}({v})\left ({\partial P\over\partial v^{\nu}}+\beta m P v^{\nu}\right )\right\rbrack,
\end{eqnarray}
where the diffusion coefficient is given by Eq. (\ref{fp3}) with
Eq. (\ref{e1}).  This equation is similar to the Kramers equation in Brownian theory (Kramers 1940) except that
the diffusion coefficient is a tensor and that it depends on the
velocity of the test star. For an isotropic distribution function (e.g., the Maxwellian), it can be put in the form
\begin{equation}
\label{e7}
D^{\mu\nu}=(D_{\|}-{1\over 2}D_{\perp}){v^{\mu}v^{\nu}\over v^{2}}+{1\over 2}D_{\perp}\delta^{\mu\nu},
\end{equation}
where $D_{\|}$ and $D_{\perp}$ are the diffusion coefficients in the directions parallel and perpendicular to the velocity of the test star. The friction by polarization can then be written as
\begin{equation}
\label{e8}
{\bf F}_{pol}=-D_{\|}
\beta m {\bf v}.
\end{equation}
The friction is proportional and opposite to the velocity of the test star, and the friction coefficient is given by the Einstein relation $\xi=D_{\|} \beta m$. The total friction is
\begin{equation}
\label{e9}
{\bf F}_{friction}=-2D_{\|} \beta m {\bf v}.
\end{equation}
This is the Chandrasekhar dynamical friction.

The steady state of the Fokker-Planck equation (\ref{e6}) is the Maxwell-Boltzmann distribution (\ref{e1}). Since the Fokker-Planck equation admits an $H$-theorem for the Boltzmann free energy (Risken 1989), one can prove that $P({\bf r},{\bf v},t)$ converges towards the Maxwell-Boltzmann distribution (\ref{e1}) for $t\rightarrow +\infty$. In other words, the test star acquires the distribution of the field stars (thermalization).  We recall, however, that for self-gravitating systems, these results cannot hold everywhere in the cluster since the Maxwell-Boltzmann distribution does not exist globally.

If we assume that the distribution $P({\bf v},t)$ of the test star is isotropic, the Fokker-Planck equation becomes
\begin{eqnarray}
\label{e10}
\frac{dP}{dt}={\partial\over\partial {\bf v}}\left\lbrack D_{\|}({v})\left ({\partial P\over\partial {\bf v}}+\beta m P {\bf v}\right )\right\rbrack,
\end{eqnarray}
It can be obtained from an effective Langevin equation 
\begin{eqnarray}
\label{e11}
\frac{d{\bf v}}{dt}=-\nabla\Phi-D_{\|}(v)\beta m {\bf v}+\frac{1}{2}\frac{\partial D_{\|}}{\partial {\bf v}}+\sqrt{2D_{\|}(v)}{\bf R}(t),
\end{eqnarray}
where ${\bf R}(t)$ is a Gaussian white noise satisfying $\langle {\bf R}(t)\rangle ={\bf 0}$ and $\langle {R}^{\mu}(t){R}^{\nu}(t')\rangle =\delta^{\mu\nu}\delta(t-t')$. Since the diffusion coefficient depends on the velocity, the noise is {\it multiplicative}. The force acting on the star consists in two parts: a part derivable from a smoothed distribution of matter $-\nabla\Phi$ and a residual random part due to the fluctuations of that distribution. In turn, the random part can be described by a Gaussian white noise multiplied by a velocity-dependent factor {\it plus} a friction force equal to the friction by polarization ${\bf F}_{pol}=-D_{\|}(v)\beta m {\bf v}$ and a spurious drift ${\bf F}_{spurious}=(1/2)({\partial D_{\|}}/{\partial {\bf v}})$ due to the multiplicative noise. If we neglect the velocity dependence of the diffusion coefficient, we recover the results of Chandrasekhar (1943a,1949) obtained in his Brownian theory.

\subsection{The diffusion tensor for isothermal systems}
\label{sec_diffusion}

When the velocity distribution of the field stars is given by the Maxwellian distribution (\ref{e1}), the diffusion tensor (\ref{fp3}) can be calculated as follows. If we introduce the representation
\begin{eqnarray}
\delta({x})=\int_{-\infty}^{+\infty} e^{i{t}{x}}\frac{d{t}}{2\pi},
\label{diff1}
\end{eqnarray}
for the $\delta$-function in Eq. (\ref{fp3}), the diffusion tensor can be rewritten as
\begin{eqnarray}
D^{\mu\nu}=\frac{1}{2}(2\pi)^{6}\int_{-\infty}^{+\infty}dt \int d{\bf k}\, k^{\mu}k^{\nu}\hat{u}(k)^{2}e^{i{\bf k}\cdot {\bf v}t} \hat{f}({\bf k}t),
\label{diff1b}
\end{eqnarray}
where $\hat{f}$ is the three-dimensional Fourier transform of the velocity distribution. This equation can be directly obtained from the formula (\ref{fpnew3b}) or (\ref{fpnew3}). This shows that the auxiliary integration variable $t$ in Eq. (\ref{diff1b}) represents the time. For the Maxwellian distribution (\ref{e1}), $\hat{f}({\bf k}t)$ is a Gaussian. If we perform the integration over  $t$ (which is the one-dimensional Fourier transform of a Gaussian), we find that the diffusion tensor can be expressed as
\begin{eqnarray}
D^{\mu\nu}=\pi (2\pi)^{3}\left (\frac{\beta m}{2\pi}\right )^{1/2}\rho m  \int \frac{k^{\mu}k^{\nu}}{k}\hat{u}(k)^{2}e^{-\beta m\frac{({\bf k}\cdot{\bf v})^{2}}{2k^{2}}}\, d{\bf k}.\nonumber\\
\label{diff2}
\end{eqnarray}
Alternatively, this expression can be obtained from Eq. (\ref{fp3}) by introducing a cartesian system of coordinates for ${\bf v}_1$ with the $z$-axis taken along the direction of ${\bf k}$, and performing the integration. With the notation $\hat{\bf k}={\bf k}/k$, Eq. (\ref{diff2}) can be rewritten as
\begin{eqnarray}
D^{\mu\nu}=\pi (2\pi)^{3}\left (\frac{\beta m}{2\pi}\right )^{1/2}\rho m  \int_{0}^{+\infty} k^{3} \hat{u}(k)^{2} d{k}\nonumber\\
\times G^{\mu\nu}\left( \sqrt{\frac{\beta m}{2}}v\right ),
\label{diff3}
\end{eqnarray}
where
\begin{eqnarray}
G^{\mu\nu}(x)=\int  \hat{k}^{\mu}\hat{k}^{\nu}e^{-(\hat{\bf k}\cdot {\bf x})^{2}}\, d{\bf \hat{k}}.
\label{diff4}
\end{eqnarray}
We note that the potential of interaction only appears in a multiplicative constant that fixes the relaxation time (see below). Using Eq. (\ref{ex1}), we get
\begin{eqnarray}
D^{\mu\nu}=\left (\frac{3}{2\pi}\right )^{1/2}\frac{2\rho m G^2\ln\Lambda}{v_m} G^{\mu\nu}\left( \sqrt{\frac{3}{2}}\frac{v}{v_m}\right ),
\label{diff5}
\end{eqnarray}
where $\ln\Lambda$ is the Coulombian factor (\ref{vl5}) and $v_{m}^{2}=3/(\beta m)$ is the mean square velocity of the field stars. The diffusion tensor may be written as
\begin{eqnarray}
\label{diff6}
D^{\mu\nu}=\frac{2v_m^2}{3t_R} G^{\mu\nu}\left( \sqrt{\frac{3}{2}}\frac{v}{v_m}\right ),
\end{eqnarray}
where $t_R$ is the local relaxation time defined below in Eq. (\ref{relax2}). This relation emphasizes the scaling $D\sim v_m^2/t_R$.

Introducing a spherical system of coordinates with the $z$-axis in the
direction of ${\bf x}$, we can write the normalized diffusion tensor
in the form
\begin{eqnarray}
G^{\mu\nu}=(G_{\|}-{1\over 2}G_{\perp}){x^{\mu}x^{\nu}\over x^{2}}+{1\over 2}G_{\perp}\delta^{\mu\nu},
\label{diff7}
\end{eqnarray}
where
\begin{eqnarray}
G_{\|}={2\pi^{3/2}\over x}G(x),\qquad
G_{\perp}={2\pi^{3/2}\over x}\lbrack {\rm erf}(x)-G(x)\rbrack,
\label{diff8}
\end{eqnarray}
with
\begin{eqnarray}
G(x)={2\over\sqrt{\pi}}{1\over x^{2}}\int_{0}^{x}t^{2}e^{-t^{2}}\, dt={1\over 2x^{2}}\biggl\lbrack {\rm erf}(x)-{2x\over \sqrt{\pi}}e^{-x^{2}}\biggr\rbrack.\nonumber\\
\label{diff9}
\end{eqnarray}
The error function is defined by
\begin{eqnarray}
{\rm erf}(x)={2\over \sqrt{\pi}}\int_{0}^{x}e^{-t^{2}}dt.
\label{diff10}
\end{eqnarray}
We have the asymptotic behaviors
\begin{eqnarray}
G_{\|}(0)=\frac{4\pi}{3}, \qquad G_{\perp}(0)=\frac{8\pi}{3},
\label{diff11}
\end{eqnarray}
\begin{eqnarray}
G_{\|}(x)\sim_{+\infty} \frac{\pi^{3/2}}{x^3}, \qquad G_{\perp}(x)\sim_{+\infty}\frac{2\pi^{3/2}}{x}.
\label{diff12}
\end{eqnarray}
We note that $G^{\mu\nu}({\bf x})\simeq G_{\|}(0)\delta^{\mu\nu}$ when 
$|{\bf x}|\rightarrow 0$.

\begin{figure}
\centering
\includegraphics[width=8cm]{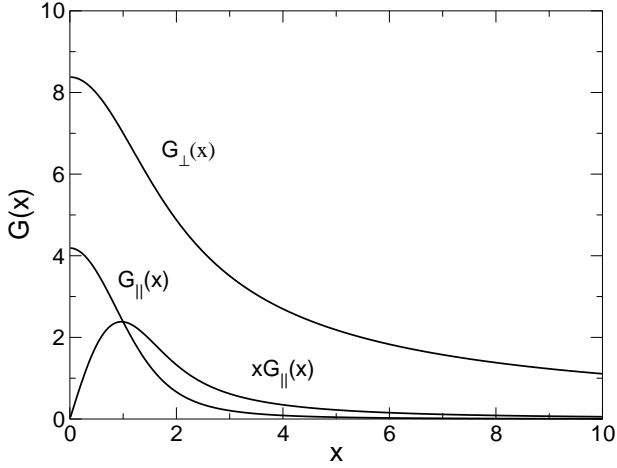}
\caption{Normalized diffusion coefficients $G_{\|}(x)$, $G_{\perp}(x)$ and friction force $xG_{\|}(x)$ for a thermal bath. The friction is maximum for $x\simeq 0.97$, i.e. when the velocity of the test star is approximately equal to the r.m.s. velocity of the field stars.}
\label{dim3}
\end{figure}

\subsection{The relaxation time}
\label{sec_relax}

We can use the preceding results to estimate the relaxation time of the velocity distribution of the 
test particle towards the Maxwellian distribution (thermalization).
If we set ${\bf
x}=\sqrt{\beta m/2}{\bf v}$, the Fokker-Planck equation
(\ref{e6}) can be rewritten as
\begin{eqnarray}
\label{relax1}
\frac{dP}{dt}=\frac{1}{t_{R}}\frac{\partial}{\partial x^{\mu}}\left\lbrack G^{\mu\nu}(x)\left (\frac{\partial P}{\partial x^{\nu}}+2 P x^{\nu}\right )\right\rbrack,
\end{eqnarray}
where $t_{R}$ is the local relaxation time
\begin{eqnarray}
\label{relax2}
{t_{R}}=\frac{1}{3}\left (\frac{2\pi}{3}\right )^{1/2}\frac{v_{m}^{3}}{\rho m G^{2}\ln\Lambda}.
\end{eqnarray}
The prefactor is equal to $0.482$ (of course this numerical factor may
vary depending on the definition of the relaxation time). The
relaxation time is inversely proportional to the local density
$\rho({\bf r})$. Therefore, the relaxation time is smaller in regions
of high density (core) and larger in regions of low density
(halo). Introducing the dynamical time $t_{D}=\lambda_J/v_{m}$, we get
\begin{eqnarray}
\label{relax3}
t_{R}=\sqrt{6\pi}\frac{n\lambda_J^3}{\ln \Lambda}t_{D}\sim \sqrt{6\pi}\frac{N}{\ln N}t_{D}.
\end{eqnarray}
We note that the relaxation time of a test particle in a bath is of the same order as the relaxation time of the system as a whole (see Sec. \ref{sec_prop}). This property is not true anymore in one dimension (Eldridge and  Feix 1962).

We can also get an estimate of the relaxation time by the following argument (Spitzer 1987).
If the diffusion coefficient were constant, the typical velocity of the test star (in one spatial direction) would increase as $\langle (\Delta {\bf v})^{2}\rangle/3\sim 2D_{\|}t$.
The relaxation time $t_{r}$ is the typical time at which the typical velocity
of the test star has reached its equilibrium value $\langle v^{2}\rangle(+\infty)=3/(m\beta)=v_{m}^{2}$ so that $\langle (\Delta {\bf v})^{2}\rangle(t_{r})=\langle v^{2}\rangle(+\infty)$. Since $D_{\|}$ depends on $v$, the description of the diffusion is more complex. However, the formula $t_{r}={v_{m}^{2}}/\lbrack {6D_{\|}(v_{m})}\rbrack$ resulting from the previous arguments with $D_{\|}=D_{\|}(v_{m})$
should provide a good estimate of the relaxation time.
Using Eq. (\ref{diff5}) and comparing with Eq. (\ref{relax2}) we obtain $t_{r}=K_{3} t_{R}$,
where $K_{3}=1/\lbrack 4G_{\|}(\sqrt{3/2})\rbrack$. Numerically, $K_{3}=0.13587547...$. 
 
Finally, we can estimate the relaxation time by $t_{r}'=\xi^{-1}$ where $\xi$ is the friction coefficient. Using the Einstein relation $\xi=2D_{\|}\beta m$ [see Eq. (\ref{e9})] with  $D_{\|}=D_{\|}(v_{m})$ we find that $t'_{r}=t_{r}$.

\subsection{The Rosenbluth potentials}
\label{sec_rosen}

It is possible to obtain simple expressions of the diffusion and friction coefficients for {\it any} isotropic distribution of the bath. If we start from the expression (\ref{vl3}) of the Vlasov-Landau equation, we find that the Fokker-Planck equation (\ref{fpnew1}) can be written as
\begin{eqnarray}
\label{rosennew1} \frac{d P}{\partial t}={\partial\over\partial
v^{\mu}}\int  K^{\mu\nu}\biggl (f_1{\partial
P\over\partial v^{\nu}}-P {\partial
f_1\over\partial v^{\nu}_1}\biggr )\, d{\bf v}_1.
\end{eqnarray}
The diffusion and friction coefficients are given by
\begin{equation}
\label{rosen1} D^{\mu\nu}=\int K^{\mu\nu} f_1 \, d{\bf v}_1=A\int f_1 \frac{w^2\delta^{\mu\nu}-w^{\mu}w^{\nu}}{w^3}\, d{\bf v}_1,
\end{equation}
\begin{eqnarray}
{F}^{\mu}_{friction}&=&2{F}^{\mu}_{pol}=2\int K^{\mu\nu} {\partial f_1\over\partial {v}_1^{\nu}} \, d{\bf v}_1\nonumber\\
&=&2\int \frac{\partial K^{\mu\nu}}{\partial v^{\nu}} f_1 \, d{\bf v}_1=-4A\int f_1 \frac{w^{\mu}}{w^3}\, d{\bf v}_1.
\label{rosen2}
\end{eqnarray}
Using the identities
\begin{eqnarray}
\label{rosen3}
K^{\mu\nu}=A\frac{\partial^{2}w}{\partial v^{\mu}\partial v^{\nu}},
\end{eqnarray}
and
\begin{eqnarray}
\label{rosen4}
\frac{\partial K^{\mu\nu}}{\partial v^{\nu}}=-2A\frac{w^{\mu}}{w^{3}}=2A  \frac{\partial}{\partial v^{\mu}}\left (\frac{1}{w}\right),
\end{eqnarray}
the coefficients of diffusion and friction can be rewritten as
\begin{eqnarray}
\label{rosen5}
D^{\mu\nu}=A \frac{\partial^{2}g}{\partial v^{\mu}\partial v^{\nu}} ({\bf v}),
\end{eqnarray}
\begin{eqnarray}
\label{rosen6}
{\bf F}_{friction}=2{\bf F}_{pol}=4 A \frac{\partial h}{\partial {\bf v}} ({\bf v}),
\end{eqnarray}
where
\begin{eqnarray}
\label{rosen7}
g({\bf v})=\int f({\bf v}_1)|{\bf v}-{\bf v}_1|d{\bf v}_1,\quad h({\bf v})=\int \frac{f({\bf v}_1)}{|{\bf v}-{\bf v}_1|}d{\bf v}_1,
\end{eqnarray}
are the so-called Rosenbluth potentials (Rosenbluth et al. 1957).

If the field particles have an isotropic velocity distribution, the
Rosenbluth potentials take the particularly simple form (see, e.g., Binney and Tremaine 2008):
\begin{eqnarray}
\label{rosen9}
h(v)=4\pi\left\lbrack \frac{1}{v}\int_{0}^{v}f(v_{1})v_{1}^{2}dv_{1}+\int_{v}^{+\infty} f(v_{1})v_{1}dv_{1}\right \rbrack,
\end{eqnarray}
\begin{eqnarray}
\label{rosen10}
g(v)=\frac{4\pi v}{3}\biggl\lbrack \int_{0}^{v}\left (3 v_{1}^{2}+\frac{v_{1}^{4}}{v^{2}}\right ) f(v_{1})dv_{1}\nonumber\\
+\int_{v}^{+\infty}\left (\frac{3v_{1}^{3}}{v}+v v_{1}\right ) f(v_{1})dv_{1}\biggr \rbrack.
\end{eqnarray}
When $g=g(v)$, the diffusion tensor (\ref{rosen5}) can be
put in the form of Eq. (\ref{e7}) with
\begin{eqnarray}
\label{rosen11}
D_{\|}=A\frac{d^{2}g}{dv^{2}},\qquad D_{\perp}=2A\frac{1}{v}\frac{dg}{dv}.
\end{eqnarray}
Using Eq. (\ref{rosen10}), we obtain
\begin{eqnarray}
\label{rosen12}
D_{\|}=\frac{8\pi}{3}A\frac{1}{v}\biggl\lbrack \int_{0}^{v}\frac{v_{1}^{4}}{v^{2}} f(v_{1})dv_{1}
+v\int_{v}^{+\infty}v_{1}f(v_{1})dv_{1}\biggr \rbrack,
\end{eqnarray}
\begin{eqnarray}
\label{rosen13}
D_{\perp}=\frac{8\pi}{3}A\frac{1}{v}\biggl\lbrack \int_{0}^{v}\left (3 v_{1}^{2}-\frac{v_{1}^{4}}{v^{2}}\right ) f(v_{1})dv_{1}\nonumber\\
+2v\int_{v}^{+\infty}v_{1}f(v_{1})dv_{1}\biggr \rbrack.
\end{eqnarray}
On the other hand, when $h=h(v)$, the friction term (\ref{rosen6}) can
be written as
\begin{eqnarray}
\label{rosen14}
{\bf F}_{friction}=2{\bf F}_{pol}=4A\frac{1}{v}\frac{dh}{dv}{\bf v}.
\end{eqnarray}
Using  Eq. (\ref{rosen9}), we get
\begin{eqnarray}
\label{rosen15}
{\bf F}_{friction}=2{\bf F}_{pol}=-16\pi A \frac{\bf v}{v^{3}} \int_{0}^{v}f(v_{1})v_{1}^{2}dv_{1}.
\end{eqnarray}
This expression can be obtained directly from Eq. (\ref{rosen6}) by noting  (Binney and Tremaine 2008) that $h({\bf v})$ in Eq. (\ref{rosen7}) is similar to the gravitational potential $\Phi({\bf r})$ produced by a distribution of mass $\rho({\bf r})$, where ${\bf v}$ plays the role of ${\bf r}$ and $f({\bf v})$ the role of $\rho({\bf r})$. Therefore, if $f(v)$ is isotropic, Eq. (\ref{rosen15}) is equivalent to the expression of the gravitational field ${\bf F}=-GM(r){\bf r}/r^3$ produced by a spherically symmetric distribution of mass,  where $M(r)$ is the mass within the sphere of radius $r$. This formula shows that the friction is due only to field stars with a velocity less than the velocity of the test star. This observation was first made by Chandrasekhar (1943a).

The previous expressions for the diffusion and the friction coefficients are valid for {\it any} isotropic
distribution of the field particles (of course, when $f({v})$ is the Maxwell distribution, we recover the results of Sec. \ref{sec_einstein}). If we substitute Eqs. (\ref{rosen12}), (\ref{rosen13}),
and (\ref{rosen14}) into Eq. (\ref{fp5}), we get a Fokker-Planck
equation describing the evolution of a test particle in a bath with a
prescribed distribution $f({v})$\footnote{Actually, this description is not self-consistent since the distribution $f({v})$ is not steady on the relaxation time $t_R$ unless it is the Maxwell distribution.}. Alternatively, if we come back to
the original Landau kinetic equation (\ref{vl3}), assume an
isotropic velocity distribution, and substitute the general expressions
(\ref{rosen12}), (\ref{rosen13}),
and (\ref{rosen14}) of the diffusion and friction coefficients with now $f=f(v,t)$ we obtain the
integro-differential equation
\begin{eqnarray}
\label{rosen16}
\frac{\partial f}{\partial t}=8\pi A \frac{1}{v^{2}}\frac{\partial}{\partial v}\biggl\lbrack \frac{1}{3}\frac{\partial f}{\partial v}\biggl ( \frac{1}{v}\int_{0}^{v}{v_{1}^{4}} f(v_{1},t)dv_{1}\nonumber\\
+v^{2}\int_{v}^{+\infty}v_{1}f(v_{1},t)dv_{1}\biggr )+f\int_{0}^{v}f(v_{1},t)v_{1}^{2}dv_{1}\biggr\rbrack,
\end{eqnarray}
describing the evolution of the system as a whole. Under this form, Eq. (\ref{rosen16}) applies to an artificial infinite homogeneous distribution of stars. This equation has been studied by King (1960) in his investigations on the evaporation of globular clusters. Within the local approximation,  Eq. (\ref{rosen16}) also represents the simplification of the collision operator that occurs in the r.h.s. of the Vlasov-Landau  equation (\ref{vl3}) when the velocity distribution of the stars is isotropic. In that case, we must restore the space variable and the advection term in Eq. (\ref{rosen16}). From this equation, implementing an adiabatic approximation, we can derive the orbit-averaged-Fokker-Planck equation which has been used by H\'enon (1961) and Cohn (1980) to study the collisional evolution of globular clusters. It reads
\begin{eqnarray}
\label{rosen17}
\frac{\partial q}{\partial \epsilon}\frac{\partial f}{\partial t}-\frac{\partial q}{\partial t}\frac{\partial f}{\partial\epsilon}=8\pi A \frac{\partial}{\partial\epsilon}\biggl\lbrack f\int_{-\infty}^{\epsilon} f_1\frac{\partial q_1}{\partial\epsilon_1}\, d\epsilon_1\nonumber\\
+\frac{\partial f}{\partial \epsilon}\left\lbrace \int_{-\infty}^{\epsilon} f_1 q_1\, d\epsilon_1+q\int_{\epsilon}^{+\infty} f_1\, d\epsilon_1\right\rbrace\bigg\rbrack,
\end{eqnarray}
where
\begin{eqnarray}
\label{rosen18}
q(\epsilon,t)=\frac{1}{3}\int_{0}^{r_{max}} \left\lbrack 2(\epsilon-\Phi(r,t))\right\rbrack^{3/2} r^2\, dr,
\end{eqnarray}
is proportional to the phase space volume available to stars with an energy less than $\epsilon$.

\subsection{Comparison with the two-body encounters theory}
\label{sec_comp}

In the previous sections, we have derived the standard kinetic equations of stellar systems from the the Liouville equation by using the BBGKY hierarchy. In standard textbooks of astrophysics (Spitzer 1987, Binney and Tremaine 2008), these equations are derived in a different manner. One usually starts from the Fokker-Planck equation (\ref{fp5}) and evaluate the diffusion tensor $\langle \Delta v^{\mu} \Delta v^{\nu}\rangle$ and the friction force $\langle \Delta v^{\mu}\rangle$ by considering the mean effect of a succession of two-body encounters. This two-body encounters theory was pioneered by Chandrasekhar (1942) and further developed by Rosenbluth et al. (1957). This approach directly leads to the expressions (\ref{rosen5}) and (\ref{rosen6}) of the diffusion and friction coefficients of a test star in a bath of field stars. These expressions are then substituted in the Fokker-Planck equation (\ref{fp5}). Finally, arguing that the field stars and the test star should evolve in the same manner, the Fokker-Planck equation is transformed into an integrodifferential equation (\ref{rosen16}) describing the evolution of the system as a whole (King 1960). When an adiabatic approximation is implemented (H\'enon 1961), one  finally obtains the orbit-averaged-Fokker-Planck equation (\ref{rosen17}).

In this paper, we have proceeded the other way round. Starting from the Liouville equation, using the BBGKY hierarchy, and making a local approximation, we have derived the Vlasov-Landau equation (\ref{vl3}) which describes the evolution of the system as a whole. Then, making a bath approximation\footnote{In the BBGKY hierarchy, this amounts to singling out a particular test star and assuming that the distribution of the other stars is fixed.}, we have obtained the Fokker-Planck equation in the form of Eq. (\ref{fp2}) with the diffusion and friction coefficients given by Eqs. (\ref{rosen1}) and (\ref{rosen2}). Our approach emphasizes the importance of the Landau equation in the kinetic theory of stellar systems, while this equation does not appear in the works of Chandrasekhar (1942), Rosenbluth et al. (1957), King (1960), and H\'enon (1961), nor in the standard textbooks of stellar dynamics by Spitzer (1987) and Binney and Tremaine (2008).

Actually,  the kinetic equation derived by these authors is equivalent to the Landau equation but it is written in a different form. They write the Fokker-Planck equation in the form of Eq. (\ref{fp5}) with the diffusion coefficient placed after the second derivative
($\partial^2 D$) while the Landau equation (\ref{vl3}) is related to the Fokker-Planck equation (\ref{fp2}) in which the diffusion coefficient
is inserted between the first derivatives
($\partial D \partial$). This difference is important on a physical point of view for two reasons. First, the Landau equation isolates the friction by polarization ${\bf F}_{pol}$ while the equation derived by Chandrasekhar (1942) and Rosenbluth et al. (1957)  involves the total friction ${\bf F}_{friction}$. Secondly, the Landau equation has a nice symmetric structure from which we can immediately deduce all the conservation laws of the system (conservation of mass, energy, impulse, and angular momentum) and the $H$-theorem for the Boltzmann entropy (Balescu 2000). These properties are less apparent in the equations derived by King (1960) and H\'enon (1961) for the evolution of the system as a whole. It is interesting to note that the symmetric structure of the kinetic equation was not realized by early stellar dynamicists while the Landau equation was known long before in plasma physics\footnote{To our knowledge, the first explicit reference to the Landau equation in the astrophysical literature appeared in the paper of Kandrup (1981).}.

Finally, the approach based on the Liouville equation and on the BBGKY
hierarchy is more rigorous than the two-body encounters theory because
it relaxes the assumption of locality and does not produce any
divergence at large scales\footnote{It is moreover applicable with
almost no modification to other systems with long-range interactions
for which a two-body encounters theory is not justified (Chavanis
2010).}. It leads to the generalized Landau equation (\ref{gen7}) that
is perfectly well-behaved at large scales contrary to the
Vlasov-Landau equation (\ref{vl3}) in which a large-scale cut-off has to be
introduced in order to avoid a divergence. This divergence is
due to the long-range nature of the gravitational potential which
precludes a rigorous application of the two-body encounters theory
that is valid for potentials with short-range interactions. Actually,
the two-body encounters theory is {\it marginally} applicable to the
gravitational force (it only generates a weak logarithmic divergence)
and this is why it is successful in practice. The generalized Landau
equation (\ref{gen7}) represents a conceptual improvement of the
Vlasov-Landau equation (\ref{vl3}) because it goes beyond the local
approximation and fully takes into account the spatial inhomogeneity
of the system. Unfortunately, this equation is very complicated to be
of much practical use. It can however be simplified by using
angle-action variables as we show in the next section.

\section{Kinetic equations with angle-action variables}
\label{sec_angleaction}

\subsection{Adiabatic approximation}
\label{sec_adiab}

In order to deal with spatially inhomogeneous systems, it is
convenient to introduce angle-action variables (Goldstein 1956, Binney and Tremaine 2008). 
Angle-action variables have
been used by many authors in
astrophysics in order to solve dynamical stability problems (Kalnajs 1977, Goodman 1988, Weinberg 1991, Pichon and Cannon 1997, Valageas 2006a) or
to compute the diffusion and friction coefficients of a test star in a cluster (Lynden-Bell and Kalnajs 1972, Tremaine and Weinberg 1984, Binney and Lacey 1988, Weinberg 1998, Nelson and Tremain 1999, Weinberg 2001, Pichon and Aubert 2006, Valageas 2006b, Chavanis 2007, Chavanis 2010). By
construction, the Hamiltonian $H$ in angle and action variables
depends only on the actions ${\bf J}=(J_1,...,J_d)$ that are constants
of the motion; the conjugate coordinates ${\bf w}=(w_1,...,w_d)$ are
called the angles (see Appendix \ref{sec_aa}). Therefore, any distribution of the form $f=f({\bf J})$
is a steady state of the Vlasov equation. According to the Jeans
theorem, this is not the general form of Vlasov steady
states. However, if the potential is regular, for all practical
purposes, any time-independent solution of the Vlasov equation may be
represented by a distribution of the form $f=f({\bf J})$ (strong Jeans
theorem).

We shall assume that the system has reached a quasi stationary
state (QSS) described by a distribution $f=f({\bf J})$ as
a result of a violent collisionless relaxation involving
only meanfield effects.  Due to finite $N$ effects, the distribution
function $f$ will slowly evolve in time. Finite $N$ effects are taken into
account in the ``collision'' operator appearing  in the
r.h.s. of Eq. (\ref{gen7}).  Since this term is of order $1/N$, the effect of
``collisions'' (granularities, finite $N$ effects, correlations...) is
a very slow process that takes place on a relaxation
timescale $t_R\sim (N/\ln N)t_{D}$ (see Sec. \ref{sec_prop}). Therefore, there is a timescale separation between the
dynamical time $t_{D}$ that is the timescale during which the system
reaches a steady state of the Vlasov equation through
phase mixing and violent collisionless relaxation and
the collisional relaxation time $t_{R}$ which is the timescale during which the system  reaches an almost isothermal  distribution due to finite $N$ effects.

Because of this timescale separation, the distribution
function is stationary on the dynamical timescale. It will 
evolve through a sequence of QSSs that are steady states  of the
Vlasov equation, depending only on the actions ${\bf J}$, slowly
changing in time due to the cumulative effect of encounters (finite $N$
effects). Indeed, the system re-adjusts itself
dynamically at each step of the collisional process. The
distribution function averaged over a short dynamical timescale can be
approximated by
\begin{equation}
\label{adiab1}
\langle f({\bf r},{\bf v},t)\rangle \simeq f({\bf J},t).
\end{equation}
Therefore, the distribution function is a function $f=f({\bf J},t)$
of the actions only that slowly evolves in time under the effect of
``collisions''.  This is similar to an adiabatic
approximation.  The system is
approximately in mechanical equilibrium at each stage of the dynamics
and the ``collisions'' slowly drive it towards an almost isothermal distribution, corresponding to a quasi thermodynamical equilibrium state.

\subsection{Evolution of the system as a whole: a Landau-type equation}
\label{sec_ses}

Introducing angle-action variables $({\bf w},{\bf J})$, the generalized Landau equation (\ref{gen7}) becomes\footnote{The adiabatic assumption (\ref{adiab1}) is consistent with the Bogoliubov ansatz of kinetic theory. Since the correlation time of the fluctuations is of the order of the dynamical time, or shorter, we can {\it freeze} the distribution function at time $t$ to compute the integral over $\tau$ (see Section \ref{sec_gen}). This distribution function, which is a steady state of the Vlasov equation,  defines a set of angle-action variables (${\bf J},{\bf w})$ that we can use to perform the integral over $\tau$. Then, the distribution function $f({\bf J},t)$ evolves with time on a longer timescale according to Eq. (\ref{ses1}).}
\begin{eqnarray}
\frac{\partial f}{\partial t}=\frac{\partial}{\partial {\bf J}}\cdot\int_{0}^{+\infty}d\tau\int d{\bf w}_1 d{\bf J}_1 {\bf F}(1\rightarrow 0)G(t,t-\tau)\nonumber\\
\times \left\lbrack {\bf F}(1\rightarrow 0)\cdot \frac{\partial}{\partial {\bf J}}+{\bf F}(0\rightarrow 1)\cdot \frac{\partial}{\partial {\bf J}_1}\right \rbrack f({\bf J},t)\frac{f}{m}({\bf J}_1,t),\quad
\label{ses1}
\end{eqnarray}
with
\begin{eqnarray}
{\bf F}(1\rightarrow 0)=-m\frac{\partial u}{\partial {\bf w}}\left\lbrack{\bf r}({\bf J},{\bf w})-{\bf r}({\bf J}_1,{\bf w}_1)\right\rbrack.
\label{ses2}
\end{eqnarray}
To obtain Eq. (\ref{ses1}), we have averaged Eq. (\ref{gen7}) over ${\bf w}$ (to simplify the expressions, the average $\langle . \rangle =(2\pi)^{-3}\int d{\bf w}$ is implicit), written the scalar products as Poisson brackets, and used the invariance of the Poisson brackets and of the phase space volume element on a change of canonical variables.
Introducing the Fourier
transform of the potential with respect to the angles
\begin{eqnarray}
A_{{\bf k},{\bf k}_1}({\bf J},{\bf J}_1)=\frac{1}{(2\pi)^{2d}}\int u\left\lbrack{\bf r}({\bf J},{\bf w})-{\bf r}({\bf J}_1,{\bf w}_1)\right\rbrack \nonumber\\
\times e^{-i({\bf k}\cdot {\bf w}-{\bf k}_1\cdot {\bf w}_1)}\, d{\bf w} d{\bf w}_1,
\label{ses4}
\end{eqnarray}
so that
\begin{eqnarray}
u\left\lbrack{\bf r}({\bf J},{\bf w})-{\bf r}({\bf J}_1,{\bf w}_1)\right\rbrack=\sum_{{\bf k},{\bf k}_1}A_{{\bf k},{\bf k}_1}({\bf J},{\bf J}_1)e^{i({\bf k}\cdot {\bf w}-{\bf k}_1\cdot {\bf w}_1)},\nonumber\\
\label{ses3}
\end{eqnarray}we get
\begin{eqnarray}
{\bf F}(1\rightarrow 0)=-im \sum_{{\bf k},{\bf k}_1}  A_{{\bf k},{\bf k}_1}({\bf J},{\bf J}_1) {\bf k} e^{i({\bf k}\cdot {\bf w}-{\bf k}_1\cdot {\bf w}_1)}.
\label{ses5}
\end{eqnarray}
Substituting this expression in Eq. (\ref{ses1}), we obtain
\begin{eqnarray}
\frac{\partial f}{\partial t}=-m^2\frac{\partial}{\partial {\bf J}}\cdot \int_{0}^{+\infty}d\tau\int d{\bf w}_1 d{\bf J_1}\sum_{{\bf k},{\bf k}_1} \sum_{{\bf l},{\bf l}_1} A_{{\bf k},{\bf k}_1}({\bf J},{\bf J}_1)\nonumber\\
\times {\bf k} e^{i({\bf k}\cdot {\bf w}-{\bf k}_1\cdot {\bf w}_1)}G(t,t-\tau)
 \biggl\lbrack A_{{\bf l},{\bf l}_1}({\bf J},{\bf J}_1) {\bf l} e^{i({\bf l}\cdot {\bf w}-{\bf l}_1\cdot {\bf w}_1)}\cdot \frac{\partial}{\partial {\bf J}}\nonumber\\
+A_{{\bf l}_1,{\bf l}}({\bf J}_1,{\bf J}) {\bf l}_1 e^{i({\bf l}_1\cdot {\bf w}_1-{\bf l}\cdot {\bf w})}\cdot \frac{\partial}{\partial {\bf J}_1}\biggr \rbrack f({\bf J},t)\frac{f}{m}({\bf J}_1,t).\nonumber\\
\label{ses6}
\end{eqnarray}
With angle-action variables, the equations of motion of a star determined by the mean field take the very simple form (see Appendix \ref{sec_aa}):
\begin{eqnarray}
{\bf J}(t-\tau)&=&{\bf J}(t)={\bf J},\nonumber\\
{\bf w}(t-\tau)&=&{\bf w}(t)-{\bf \Omega}({\bf J},t)\tau={\bf w}-{\bf \Omega}({\bf J},t)\tau,
\label{ses7}
\end{eqnarray}
where ${\bf \Omega}({\bf J},t)$ is the angular frequency of the orbit with action
${\bf J}$. As explained previously, we have neglected the variation of the mean field on a timescale of the order of the dynamical time so it is considered as ``frozen'' when we compute the stellar trajectories  (adiabatic or Bogoliubov assumption). Substituting these relations in Eq. (\ref{ses6})
and making the transformations ${\bf l}\rightarrow -{\bf l}$ and ${\bf l}_1\rightarrow
-{\bf l}_1$ in the second term (friction term), we obtain successively
\begin{eqnarray}
\frac{\partial f}{\partial t}=-m^2\frac{\partial}{\partial {\bf J}}\cdot \int_{0}^{+\infty}d\tau\int d{\bf w}_1 d{\bf J}_1 \sum_{{\bf k},{\bf k}_1} \sum_{{\bf l},{\bf l}_1} A_{{\bf k},{\bf k}_1}({\bf J},{\bf J}_1) \nonumber\\
\times{\bf k} e^{i({\bf k}\cdot {\bf w}-{\bf k}_1\cdot {\bf w}_1)}
 e^{i({\bf l}\cdot {\bf w}(t-\tau)-{\bf l}_1\cdot {\bf w}_1(t-\tau))}\nonumber\\
 \times\left\lbrack A_{{\bf l},{\bf l}_1}({\bf J},{\bf J}_1) {\bf l}\cdot  \frac{\partial}{\partial {\bf J}}-A_{-{\bf l}_1,-{\bf l}}({\bf J}_1,{\bf J}) {\bf l}_1 \cdot \frac{\partial}{\partial {\bf J}_1}\right \rbrack \nonumber\\
 \times f({\bf J},t)\frac{f}{m}({\bf J}_1,t),\qquad
\label{ses8}
\end{eqnarray}
and
\begin{eqnarray}
\frac{\partial f}{\partial t}=-m^2\frac{\partial}{\partial {\bf J}}\cdot \int_{0}^{+\infty}d\tau\int d{\bf w}_1 d{\bf J}_1 \sum_{{\bf k},{\bf k}_1} \sum_{{\bf l},{\bf l}_1} A_{{\bf k},{\bf k}_1}({\bf J},{\bf J}_1) \nonumber\\
\times{\bf k} e^{i({\bf k}\cdot {\bf w}-{\bf k}_1\cdot {\bf w}_1)}
 e^{i({\bf l}\cdot {\bf w}-{\bf l}_1\cdot {\bf w}_1)}e^{-i({\bf l}\cdot {\bf \Omega}({\bf J},t)-{\bf l}_1\cdot {\bf \Omega}({\bf J}_1,t))\tau}\nonumber\\
\times \left\lbrack A_{{\bf l},{\bf l}_1}({\bf J},{\bf J}_1) {\bf l}\cdot  \frac{\partial}{\partial {\bf J}}-A_{-{\bf l}_1,-{\bf l}}({\bf J}_1,{\bf J}) {\bf l}_1 \cdot \frac{\partial}{\partial {\bf J}_1}\right \rbrack \nonumber\\
\times f({\bf J},t)\frac{f}{m}({\bf J}_1,t).\qquad
\label{ses9}
\end{eqnarray}
It is easy to establish that
\begin{eqnarray}
A_{{\bf k}_1,{\bf k}}({\bf J}_1,{\bf J})=A_{-{\bf k},-{\bf k}_1}({\bf J},{\bf J}_1)=A_{{\bf k},{\bf k}_1}({\bf J},{\bf J}_1)^*.
\label{ses10}
\end{eqnarray}
Therefore, the kinetic equation can be rewritten as
\begin{eqnarray}
\frac{\partial f}{\partial t}=-m^2\frac{\partial}{\partial {\bf J}}\cdot \int_{0}^{+\infty}d\tau\int d{\bf w}_1 d{\bf J}_1 \sum_{{\bf k},{\bf k}_1} \sum_{{\bf l},{\bf l}_1} A_{{\bf k},{\bf k}_1}({\bf J},{\bf J}_1) \nonumber\\
\times{\bf k} e^{i({\bf k}+{\bf l})\cdot {\bf w}}e^{-i({\bf k}_1+{\bf l}_1)\cdot {\bf w}_1}
 e^{-i({\bf l}\cdot {\bf \Omega}({\bf J},t)-{\bf l}_1\cdot {\bf \Omega}({\bf J}_1,t))\tau}\nonumber\\
 A_{{\bf l},{\bf l}_1}({\bf J},{\bf J}_1)\left (  {\bf l}\cdot  \frac{\partial}{\partial {\bf J}}- {\bf l}_1 \cdot \frac{\partial}{\partial {\bf J}_1}\right ) f({\bf J},t)\frac{f}{m}({\bf J}_1,t).\qquad
\label{ses11}
\end{eqnarray}
Integrating over ${\bf w}_1$, and recalling that this expression has to be averaged over ${\bf w}$, we obtain
\begin{eqnarray}
\frac{\partial f}{\partial t}=(2\pi)^3 m^2\frac{\partial}{\partial {\bf J}}\cdot \int_{0}^{+\infty}d\tau\int d{\bf J}_1 \sum_{{\bf k},{\bf k}_1} |A_{{\bf k},{\bf k}_1}({\bf J},{\bf J}_1)|^2  \nonumber\\
\times {\bf k}
 e^{-i(-{\bf k}\cdot {\bf \Omega}({\bf J},t)+{\bf k}_1\cdot {\bf \Omega}({\bf J}_1,t))\tau}\nonumber\\
 \times
 \left (  {\bf k}\cdot  \frac{\partial}{\partial {\bf J}}-{\bf k}_1 \cdot \frac{\partial}{\partial {\bf J}_1}\right ) f({\bf J},t)\frac{f}{m}({\bf J}_1,t).\qquad
\label{ses12}
\end{eqnarray}
Making the transformation $\tau\rightarrow -\tau$, then $({\bf k},{\bf k}_1)\rightarrow (-{\bf k},-{\bf k}_1)$, and adding the resulting expression to Eq. (\ref{ses12}), we get
\begin{eqnarray}
\frac{\partial f}{\partial t}=\frac{1}{2}(2\pi)^3 m^2\frac{\partial}{\partial {\bf J}}\cdot \int_{-\infty}^{+\infty}d\tau\int d{\bf J}_1 \sum_{{\bf k},{\bf k}_1} |A_{{\bf k},{\bf k}_1}({\bf J},{\bf J}_1)|^2  \nonumber\\
\times {\bf k}
 e^{i({\bf k}\cdot {\bf \Omega}({\bf J},t)-{\bf k}_1\cdot {\bf \Omega}({\bf J}_1,t))\tau}\nonumber\\
 \times
 \left (  {\bf k}\cdot  \frac{\partial}{\partial {\bf J}}-{\bf k}_1 \cdot \frac{\partial}{\partial {\bf J}_1}\right ) f({\bf J},t)\frac{f}{m}({\bf J}_1,t).\qquad
\label{ses13}
\end{eqnarray}
Finally, using the identity (\ref{sw13}), we obtain the kinetic equation 
\begin{eqnarray}
\frac{\partial f}{\partial t}=\pi(2\pi)^3 m\frac{\partial}{\partial {\bf J}}\cdot \sum_{{\bf k},{\bf k}_1}\int d{\bf J}_1 {\bf k} |A_{{\bf k},{\bf k}_1}({\bf J},{\bf J}_1)|^2  \nonumber\\
\times \delta\left\lbrack {\bf k}\cdot {\bf \Omega}({\bf J},t)-{\bf k}_1\cdot {\bf \Omega}({\bf J}_1,t)\right\rbrack\nonumber\\
 \times
 \left (  {\bf k}\cdot  \frac{\partial}{\partial {\bf J}}-{\bf k}_1 \cdot \frac{\partial}{\partial {\bf J}_1}\right ) f({\bf J},t)f({\bf J}_1,t).\qquad
\label{ses14}
\end{eqnarray}
This kinetic equation  was previously derived for systems with arbitrary long-range interactions in various dimensions of space (Chavanis 2007,2010) and it is here specifically applied to stellar systems. Since collective effects are neglected, this kinetic equation can be viewed as a Landau-type equation with angle-action variables describing the evolution of spatially inhomogeneous stellar  systems. The collisional evolution of these systems is due to a condition of resonance $ {\bf k}\cdot {\bf \Omega}({\bf J},t)={\bf k}_1\cdot {\bf \Omega}({\bf J}_1,t)$ (with $({\bf k}_1,{\bf J}_1)\neq ({\bf k},{\bf J})$) encapsulated in the $\delta$-function. This $\delta$-function expresses the conservation of energy. It can be shown (Chavanis 2007) that the kinetic equation (\ref{ses14}) conserves mass $M=\int f\, d{\bf J}$ and energy $E=\int f\epsilon({\bf J})\, d{\bf J}$ and monotonically increases the Boltzmann entropy $S=-\int (f/m)\ln (f/m)\, d{\bf J}$ ($H$-theorem). However, as explained in Section \ref{sec_prop}, this equation does not reach a steady state due to the absence of statistical equilibrium for stellar systems\footnote{For self-gravitating systems in lower dimensions of space, or for systems with long-range interactions with a smooth potential like the HMF model, a statistical equilibrium state exists. In that case, it can be shown that the Landau-type equation (\ref{ses14}) relaxes towards the Boltzmann distribution on a timescale $Nt_D$  provided there are enough resonances (see Chavanis 2007).}.

\subsection{Relaxation of a star in a thermal bath: the Fokker-Planck equation}
\label{sec_ta}

Implementing a test particle approach as in Sec. \ref{sec_tp}, we find that the equation for $P({\bf J},t)$, the probability density of finding the test star with an action ${\bf J}$ at time $t$, is
\begin{eqnarray}
\frac{\partial P}{\partial t}=\pi(2\pi)^3 m\frac{\partial}{\partial {\bf J}}\cdot \sum_{{\bf k},{\bf k}_1}\int d{\bf J}_1 {\bf k} |A_{{\bf k},{\bf k}_1}({\bf J},{\bf J}_1)|^2  \nonumber\\
\times \delta \lbrack {\bf k}\cdot {\bf \Omega}({\bf J})-{\bf k}_1\cdot {\bf \Omega}({\bf J}_1)\rbrack \nonumber\\
 \times
 \left (  {\bf k}\cdot  \frac{\partial}{\partial {\bf J}}-{\bf k}_1 \cdot \frac{\partial}{\partial {\bf J}_1}\right ) P({\bf J},t)f({\bf J}_1).\qquad
\label{ta1}
\end{eqnarray}
The angular frequency  ${\bf \Omega}({\bf J})$ is now a {\it static} function determined by the distribution $f({\bf J})$ of the field stars. Equation (\ref{ta1}) can be written in the form of a Fokker-Planck equation
\begin{equation}
\label{ta2}{\partial P\over\partial t}={\partial\over\partial {J}^{\mu}} \biggl (D^{\mu\nu}{\partial P\over\partial {J}^{\nu}}-P {F}_{pol}^{\mu}\biggr ),
\end{equation}
involving a diffusion tensor
\begin{eqnarray}
D^{\mu\nu}=\pi(2\pi)^3 m\sum_{{\bf k},{\bf k}_1}\int d{\bf J}_1\,  {k}^{\mu} k^{\nu}  |A_{{\bf k},{\bf k}_1}({\bf J},{\bf J}_1)|^2  \nonumber\\
\times \delta\lbrack{\bf k}\cdot {\bf \Omega}({\bf J})-{\bf k}_1\cdot {\bf \Omega}({\bf J}_1)\rbrack f({\bf J}_1),\qquad
\label{ta3}
\end{eqnarray}
and a friction by polarization
\begin{eqnarray}
{\bf F}_{pol}=\pi(2\pi)^3 m\sum_{{\bf k},{\bf k}_1}\int d{\bf J}_1\, {\bf k} |A_{{\bf k},{\bf k}_1}({\bf J},{\bf J}_1)|^2  \nonumber\\
\times \delta\lbrack{\bf k}\cdot {\bf \Omega}({\bf J})-{\bf k}_1\cdot {\bf \Omega}({\bf J}_1)\rbrack {\bf k}_1\cdot \frac{\partial f}{\partial {\bf J}_1}({\bf J}_1).\qquad
\label{ta4}
\end{eqnarray}
Writing the Fokker-Planck equation in the usual form
\begin{equation}
\label{ta5} {\partial P\over\partial t}={\partial^2\over\partial J^{\mu}\partial J^{\nu}}(D^{\mu\nu}P)-{\partial\over\partial J_i}(P F_{friction}^{\mu}),
\end{equation}
with
\begin{equation}
\label{ta6}
D^{\mu\nu}={\langle \Delta J^{\mu} \Delta J^{\nu}\rangle\over 2 \Delta t}, \quad {\bf F}_{friction}={\langle \Delta {\bf J}\rangle\over \Delta t},
\end{equation}
we find that the relation between the friction by polarization and the total friction is
\begin{equation}
\label{ta6b}
{F}^{\mu}_{friction}={F}^{\mu}_{pol}+\frac{\partial D^{\mu\nu}}{\partial
J^{\nu}}.
\end{equation}
Substituting Eqs. (\ref{ta3}) and (\ref{ta4}) in Eq. (\ref{ta6b}) and using an integration by parts, we find that the diffusion and friction coefficients are given by
\begin{eqnarray}
\label{ta7}{\langle \Delta J^{\mu} \Delta J^{\nu}\rangle\over 2 \Delta t}=\pi(2\pi)^3 m\int d{\bf J}_1\,  f({\bf J}_1) \sum_{{\bf k},{\bf k}_1}  {k}^{\mu} k^{\nu}   \nonumber\\
\times |A_{{\bf k},{\bf k}_1}({\bf J},{\bf J}_1)|^2 \delta({\bf k}\cdot {\bf \Omega}({\bf J})-{\bf k}_1\cdot {\bf \Omega}({\bf J}_1)),
\end{eqnarray}
\begin{eqnarray}
{\langle \Delta {\bf J}\rangle\over \Delta t}=\pi(2\pi)^3 m\int d{\bf J}_1\, f({\bf J}_1)\sum_{{\bf k},{\bf k}_1}{\bf k}  \left ({\bf k}\cdot \frac{\partial}{\partial {\bf J}}-{\bf k}_1\cdot \frac{\partial}{\partial {\bf J}_1}\right )\nonumber\\
\times |A_{{\bf k},{\bf k}_1}({\bf J},{\bf J}_1)|^2 \delta({\bf k}\cdot {\bf \Omega}({\bf J})-{\bf k}_1\cdot {\bf \Omega}({\bf J}_1)).\qquad
\label{ta8}
\end{eqnarray}
These expressions can be obtained directly from the Hamiltonian
equations of motion by expanding the trajectories of the stars
in powers of $1/N$ in the limit $N\rightarrow +\infty$ (Valageas 2006a).

Let us assume that the field stars form a thermal bath with the Boltzmann distribution
\begin{eqnarray}
f({\bf J}_1)=A e^{-\beta m\epsilon({\bf J}_1)},\label{ta9}
\end{eqnarray}
where $\epsilon({\bf J})$ is the energy of a star in an orbit with action ${\bf J}$. As we have explained before, this distribution is not defined globally for a self-gravitating system. However, it holds approximately for stars with low energy\footnote{This description is also valid for self-gravitating systems in lower dimension of space, or for other systems with long-range interaction, for which a statistical equilibrium state exists (Chavanis 2007, 2010).}. Using the identity $\partial\epsilon/\partial{\bf J}={\bf \Omega}({\bf J})$ (see Appendix \ref{sec_aa}), we find that
\begin{eqnarray}
\frac{\partial f_1}{\partial{\bf J}_1}=-\beta m f({\bf J}_1) {\bf \Omega}({\bf J}_1).
\label{ta10}
\end{eqnarray}
Substituting this relation in Eq. (\ref{ta4}), using the $\delta$-function to replace ${\bf k}_1\cdot {\bf \Omega}({\bf J}_1)$ by ${\bf k}\cdot {\bf \Omega}({\bf J})$, and  comparing the resulting
expression with Eq. (\ref{ta3}), we finally get
\begin{eqnarray}
F^{\mu}_{pol}=-D^{\mu\nu}({\bf J})\beta m {\Omega}^{\nu}({\bf J}), \label{ta12}
\end{eqnarray}
which is the appropriate Einstein relation for our problem.
For a thermal bath, using Eq. (\ref{ta12}), the Fokker-Planck equation
(\ref{ta2}) can be written as
\begin{equation}
\label{ta13}{\partial P\over\partial
t}={\partial\over\partial J^{\mu}}\biggl\lbrack  D^{\mu\nu}({\bf J})\biggl
({\partial P\over\partial J^{\nu}}+\beta m P\Omega^{\nu}({\bf J})\biggr
)\biggr\rbrack,
\end{equation}
where $D^{\mu\nu}({\bf J})$ is given by Eq. (\ref{ta3}) with Eq. (\ref{ta9}). Recalling that ${\bf\Omega}({\bf
J})=\partial \epsilon/\partial {\bf J}$, this equation is similar to
the Kramers equation in Brownian theory (Kramers 1940). This is a
drift-diffusion equation describing the evolution of the distribution
$P({\bf J},t)$ of the test star in an ``effective potential''
$U_{eff}({\bf J})=\epsilon({\bf J})$ produced by the field stars.
For $t\rightarrow +\infty$, the distribution of the test
star relaxes towards the Boltzmann distribution (\ref{ta9}). This takes place on a typical
relaxation time $t_R\sim (N/\ln N)t_D$. Again, this is valid only in the part of the cluster where the Boltzmann distribution holds approximately.

\section{Conclusion}
\label{sec_conclusion}

Starting from the Liouville equation, using a truncation of the BBGKY hierarchy at the order $1/N$,  and neglecting collective effects, we have derived a kinetic equation (\ref{gen7}) in physical space that can be viewed as a generalized Landau equation. This equation was previously derived by Kandrup  (1981) using projection operator technics. A nice feature of this equation is that it does not present any divergence at large scales since the spatial inhomogeneity of the system is accounted for.  When a local approximation is implemented, and a cut-off is introduced heuristically at the Jeans length, we recover the Vlasov-Landau equation (\ref{vl2}) which is the standard equation  of stellar dynamics. On the other hand, using angle-action variables,  we have derived a Landau-type equation (\ref{ses14}) for fully inhomogeneous stellar systems. We have also developed a test particle approach and derived the corresponding Fokker-Planck equations (\ref{fp1}) and (\ref{ta1}) in position-velocity space and angle-action space respectively. Explicit expressions have been given for the diffusion and friction coefficients. We have distinguished the friction by polarization from the total friction. A limitation of the approach presented here is that it neglects collective effects. More general kinetic equations,  corresponding to Lenard-Balescu-type equations taking spatial inhomogeneity and collective effects into account, have been derived recently by Heyvaerts (2010) from the Liouville equation and by Chavanis (2012b) from the Klimontovich equation (these approaches based on the BBGKY hierarchy or on the quasilinear approximation are equivalent but the formalism is different). These kinetic equations are  more general than those derived in the present paper, but they are also more complicated (to derive and to solve). Therefore, the equations presented in this paper may be useful as a first step.

We have also discussed the differences between the present approach based on the BBGKY hierarchy and the more classical two-body encounters theory (Chandrasekhar 1942).  The two-body encounters theory, which is usually adapted to short-range potentials (Boltzmann 1872, Chapman and Cowling 1939), can take strong collisions into account so it does not yield any divergence at small scales. However, this approach cannot take spatial inhomogeneity into account so it leads to a divergence at large scales. This divergence is due to the long-range nature of the gravitational potential and the dominance of weak collisions. In addition, the two-body encounters theory does not take collective effects into account; these effects are specific to systems with long-range interactions. By contrast, the approach based on the BBGKY hierarchy takes into account the spatial inhomogeneity of the system and collective effects. Therefore, it does not yield any divergence at large scales. However, it fails to take strong collisions into account due to the weak coupling approximation. In a sense, the gravitational potential is intermediate between short-range and long-range potentials because both strong collisions and weak collisions are relevant. Therefore, the approaches adapted to short-range or long-range potentials are both {\it marginally} applicable (they yield a logarithmic divergence are large or small scales respectively). This is why the kinetic equations of stellar dynamics can be obtained in different manners that turn out to be complementary to each other.

In this paper, we have assumed that the system is isolated from the surrounding. As a result, the source of noise is due to discreteness (finite $N$) effects internal to the system. The case where the noise is caused by external sources (perturbations on a galaxy, cosmological environment on dark matter halos...) is also interesting. It has been considered by several authors such as Weinberg (2001), Ma and Bertschinger (2004), and Pichon and Aubert (2006) who developed appropriate kinetic theories.

\appendix

\section{The thermodynamic limit}
\label{sec_tl}

The kinetic and potential energies in the Hamiltonian (\ref{bbgky1})
are comparable provided that $v_m^2\sim GNm/R$, where $v_m$ is the
root mean square velocity of the stars and $R$ the system's size (this
scaling can also be obtained from the virial theorem). Therefore, the
energy scales as $E\sim Nmv_m^2\sim GN^2m^2/R$ and the kinetic
temperature, defined by $k_B T=m v_m^2/3$, scales as $k_BT\sim
GNm^2/R\sim E/N$. The thermodynamic limit of a self-gravitating system
corresponds to $N\rightarrow +\infty$ in such a way that the
normalized energy $\epsilon=ER/(GN^2m^2)$ and the normalized
temperature $\eta=\beta GNm^2/R$ are of order unity. Of course, the
usual thermodynamic limit $N,V\rightarrow +\infty$ with $N/V\sim 1$ is
not applicable to self-gravitating systems since these systems are
spatially inhomogeneous.

By a suitable normalization of the parameters, we can take $R\sim 1$,
$m\sim 1$, and $G\sim 1/N$. In this way, $E\sim N$, $S\sim N$ and
$T\sim 1$. This is the proper thermodynamic limit for systems with
long-range interactions (Kac et al. 1963, Campa et al. 2009). We note
that the coupling constant $G$ scales as $1/N$. The energy and the
entropy are extensive but they remain fundamentally non-additive. The
temperature is intensive. The dynamical time $t_D\sim R/v_m\sim
1/\sqrt{G\rho}$ is of order unity ($t_D\sim 1$).

Other normalizations of the parameters are possible. For example,
Gilbert (1968) considers the limit $N\rightarrow +\infty$ with $G\sim
1$, $R\sim 1$, and $m\sim 1/N$. In that case, $E\sim 1$, $S\sim N$,
$T\sim 1/N$, and $t_D\sim 1$. On the other hand, de Vega and Sanchez
(2002) define the thermodynamic limit as $N\rightarrow +\infty$ with
$m\sim 1$, $G\sim 1$, and $R\sim N$. In that case, $E\sim N$, $S\sim
N$, and $T\sim 1$. However, the dynamical time $t_D\sim N$ diverges
with the number of particles. Therefore, this normalization may not be
convenient to develop the kinetic theory of stellar systems. If we
impose $G\sim 1$, $E\sim N$, $T\sim 1$ {\it and} $t_D\sim 1$, we get
$R\sim N^{1/5}$ and $m\sim N^{-2/5}$.

Dimensionally, the Jeans length scales as $\lambda_J\sim (k_BT/Gm^2
n)^{1/2}$, where $n$ is the number density. This is the counterpart of
the Debye length $\lambda_D\sim (k_B T/ne^2)^{1/2}$ in plasma
physics. Since $v_m^2\sim GM/R$, we find that $R\sim
{v_m}/({Gnm})^{1/2}\sim ({k_BT}/{Gnm^2})^{1/2}\sim
\lambda_J$. Therefore, the Jeans length is of the order of the
system's size.

The gravitational parameter (resp. the plasma parameter) is defined as
the ratio of the interaction strength at the mean interparticle
distance $Gm^2n^{1/3}$ (resp. $e^2n^{1/3}$) to the thermal energy $k_B
T$. This leads to $g=G m^2 n^{1/3}/k_B
T=1/(n\lambda_J^3)^{2/3}=1/\Lambda^{2/3}\sim 1/N^{2/3}$ (resp. $g=e^2
n^{1/3}/k_B T=1/(n\lambda_D^3)^{2/3}=1/\Lambda^{2/3}$). Estimating the
density by $n\sim N/R^3$, we find that $g\sim
\eta/N^{2/3}$. Therefore, the expansion of the BBGKY hierarchy in
terms of the gravitational parameter (resp. plasma parameter)
$g$ is equivalent to an expansion in terms of the inverse of the
number of particles in the Jeans sphere $\Lambda=n\lambda_J^3\sim N$
(resp. the inverse of the number of particles in the Debye sphere
$\Lambda=n\lambda_D^3$). This corresponds to a weak coupling
approximation.

\section{Angle-action variables} 
\label{sec_aa}

In Section \ref{sec_angleaction}, we have explained that, during its collisional evolution, a stellar system passes by a succession of QSSs that are steady states of the Vlasov equation slowly changing under the effect of close encounters (finite $N$ effects). The slowly varying distribution function $f({\bf r},{\bf v})$
determines a potential $\Phi({\bf r})$ and a one-particle Hamiltonian $\epsilon=v^2/2+\Phi({\bf r})$ that, we assume,
is integrable. Therefore, it is possible to use  angle-action variables constructed with this Hamiltonian (Goldstein 1956, Binney and Tremaine 2008). This construction is done adiabatically, i.e. the distribution function, and the  angle-action variables, slowly change in time.

A particle with coordinates $({\bf r},{\bf v})$ in
phase space is described equivalently by the angle-action
variables $({\bf w},{\bf J})$. The Hamiltonian equations for the
conjugate variables $({\bf r},{\bf v})$ are
\begin{equation}
\label{aa1}
\frac{d{\bf r}}{dt}=\frac{\partial \epsilon}{\partial {\bf v}}={\bf v}, \qquad \frac{d{\bf v}}{dt}=-\frac{\partial \epsilon}{\partial {\bf r}}=-\nabla\Phi({\bf r}).
\end{equation}
In terms of the variables $({\bf r},{\bf v})$, the dynamics
is complicated because the potential explicitly appears in
the second equation. Therefore, this equation $d{\bf
v}/dt=-\nabla\Phi$ cannot be easily integrated except if $\Phi=0$,
i.e. for a spatially homogeneous system. In that case, the velocity
${\bf v}$ is constant and the unperturbed equations of motion reduce
to ${\bf r}={\bf v}t+{\bf r}_{0}$, i.e. to a rectilinear motion at
constant velocity.  Now, the angle-action variables are constructed so
that the Hamiltonian does not depend on the angles ${\bf
w}$. Therefore, the Hamiltonian equations for the conjugate
variables $({\bf w},{\bf J})$ are
\begin{equation}
\label{aa2}
\frac{d{\bf w}}{dt}=\frac{\partial \epsilon}{\partial {\bf J}}= {\bf \Omega}({\bf J}), \qquad \frac{d{\bf J}}{dt}=-\frac{\partial \epsilon}{\partial {\bf w}}={\bf 0},
\end{equation}
where ${\bf \Omega}({\bf J})$ is
the angular frequency of the orbit with action ${\bf J}$. From these equations, we find
that ${\bf J}$ is a constant and that ${\bf w}={\bf \Omega}({\bf J})t+{\bf
w}_{0}$. Therefore, the equations of
motion are very simple in these variables. They  extend naturally the
trajectories at constant velocity for spatially homogeneous systems. This is
why this choice of variables is relevant to develop the kinetic
theory. Of course, even if the description of the motion becomes simple in these variables,
the complexity of the problem has not completely disappeared. It is now embodied in the relation between position and momentum variables and angle and action variables which can be quite complicated.

\section{Calculation of $K^{\mu\nu}$}
\label{sec_sw}

Within the local approximation, we can proceed as if the system were spatially homogeneous. In that case, the  mean field force vanishes, $\langle {\bf F}\rangle={\bf 0}$, and the unperturbed equations of motion (i.e. for $N\rightarrow +\infty$) reduce to
\begin{eqnarray}
{\bf v}(t-\tau)={\bf v}(t)={\bf v},
\label{sw5a}
\end{eqnarray}
\begin{eqnarray}
{\bf r}(t-\tau)={\bf r}(t)-{\bf v}(t)\tau={\bf r}-{\bf v}\tau,
\label{sw5b}
\end{eqnarray}
corresponding to a rectilinear motion at constant velocity.  The collision term in the kinetic equation (\ref{vl1}) can be written as
\begin{eqnarray}
\left (\frac{\partial f}{\partial t}\right )_{coll}=\frac{\partial}{\partial
{v}^{\mu}}\int d{\bf v}_1 K^{\mu\nu}
 \left ({\partial\over\partial { v}^{\nu}}-{\partial\over\partial {v}_{1}^{\nu}}\right )\nonumber\\
 \times {f}({\bf r},{\bf v},t)f({\bf r},{\bf
v}_1,t), \label{sw6}
\end{eqnarray}
with
\begin{eqnarray}
K^{\mu\nu}=\frac{1}{m}\int_0^{+\infty} d\tau \int d{\bf r}_{1}
{F}^{\mu}(1\rightarrow
0,t){{F}}^{\nu}(1\rightarrow 0,t-\tau).\nonumber\\ \label{sw7}
\end{eqnarray}
The force by unit of mass created by particle $1$ on particle $0$ is given by
\begin{eqnarray}
{\bf F}(1\rightarrow 0)=-m\frac{\partial u}{\partial {\bf r}}({\bf r}-{\bf r}_{1}),\label{sw8}
\end{eqnarray}
where $u({\bf r}-{\bf r}')=-G/|{\bf r}-{\bf r}'|$ is the gravitational potential. The Fourier transform, and the inverse Fourier transform, of the potential are defined by
\begin{eqnarray}
\hat{u}({\bf k})=\int e^{-i{\bf k}\cdot {\bf x}}{u}({\bf x})\, \frac{d{\bf x}}{(2\pi)^d},\qquad u({\bf x})=\int e^{i{\bf k}\cdot {\bf x}}\hat{u}({\bf k})\, d{\bf k}.\nonumber\\
\label{sw9}
\end{eqnarray}
For the gravitational interaction
\begin{eqnarray}
\label{ex1}
(2\pi)^{3}\hat{u}(k)=-\frac{4\pi G}{k^{2}}.
\end{eqnarray}
Substituting Eq. (\ref{sw9}-b) in Eq. (\ref{sw8}), and writing explicitly the Lagrangian coordinates, we get
\begin{eqnarray}
{\bf F}(1\rightarrow 0,t-\tau)=-i m\int {\bf k} \, e^{i{\bf k}\cdot ({\bf r}(t-\tau)-{\bf r}_{1}(t-\tau))}\hat{u}({\bf k})\, d{\bf k}.\nonumber\\ \label{sw10}
\end{eqnarray}
Using the equations of motion (\ref{sw5a}) and (\ref{sw5b}), and introducing the notations ${\bf x}={\bf r}-{\bf r}_{1}$ and ${\bf w}={\bf v}-{\bf v}_{1}$, we obtain
\begin{eqnarray}
{\bf F}(1\rightarrow 0,t-\tau)
=-i m\int {\bf k} \, e^{i{\bf k}\cdot ({\bf x}-{\bf w}\tau)}\hat{u}({\bf k})\, d{\bf k}.\label{sw11}
\end{eqnarray}
Therefore
\begin{eqnarray}
K^{\mu\nu}=-m \int_{0}^{+\infty}d\tau\int d{\bf x} \int d{\bf k} \int d{\bf k}' k_{\mu} k'_{\nu}\nonumber\\
\times e^{i({\bf k}+{\bf k}')\cdot {\bf x}} e^{-i{\bf k}'\cdot {\bf w}\tau}\hat{u}({\bf k})\hat{u}({\bf k}').\label{sw12}
\end{eqnarray}
Using the identity
\begin{eqnarray}
\delta({\bf x})=\int e^{i{\bf k}\cdot {\bf x}}\, \frac{d{\bf k}}{(2\pi)^d}, \label{sw13}
\end{eqnarray}
and integrating over ${\bf x}$ and ${\bf k}'$, we find that
\begin{eqnarray}
K^{\mu\nu}=(2\pi)^3 m  \int_{0}^{+\infty}d\tau \int d{\bf k} \,  k^{\mu} k^{\nu}  e^{i{\bf k}\cdot {\bf w}\tau}\hat{u}({k})^2.\label{sw15}
\end{eqnarray}
Performing the transformation $\tau\rightarrow -\tau$, then ${\bf k}\rightarrow -{\bf k}$, and adding the resulting expression to Eq. (\ref{sw15}), we get
\begin{eqnarray}
K^{\mu\nu}=\frac{1}{2}(2\pi)^3 m  \int_{-\infty}^{+\infty}d\tau \int d{\bf k}  \, k^{\mu} k^{\nu} e^{i{\bf k}\cdot {\bf w}\tau}\hat{u}({k})^2.\label{sw16}
\end{eqnarray}
Using the identity (\ref{sw13}), we finally obtain
\begin{eqnarray}
K^{\mu\nu}=\pi (2\pi)^3 m \int d{\bf k} \,  k^{\mu}k^{\nu} \delta({\bf k}\cdot {\bf w}) \hat{u}({k})^2,\label{sw17}
\end{eqnarray}
which leads to Eq. (\ref{vl2}).

Introducing a spherical system of
coordinates in which the $z$ axis is taken in the direction of ${\bf
w}$, we obtain
\begin{eqnarray}
K^{\mu\nu}=\pi (2\pi)^3 m \int_{0}^{+\infty} k^2dk\int_{0}^{\pi}\sin\theta\, d\theta \nonumber\\
\times \int_{0}^{2\pi}d\phi\,   k^{\mu} k^{\nu} \delta(kw\cos\theta) \hat{u}(k)^2.\label{k1}
\end{eqnarray}
Using $k_{x}=k\sin\theta\cos\phi$, $k_y=k\sin\theta\sin\phi$ and
$k_z=k\cos\theta$, it is easy to see that only $K_{xx}$, $K_{yy}$ and
$K_{zz}$ can be non-zero. The other components of the matrix
$K^{\mu\nu}$ vanish by symmetry. Furthermore
\begin{eqnarray}
K_{xx}=K_{yy}=\pi^2 (2\pi)^3 m \int_{0}^{+\infty} k^2\, dk\nonumber\\
\times\int_{0}^{\pi}\sin\theta\, d\theta\,   k^{2} \delta(kw\cos\theta) \hat{u}(k)^2 \sin^{2}\theta.
\label{k2}
\end{eqnarray}
Using the identity $\delta(\lambda x)=\frac{1}{|\lambda|}\delta(x)$, we get
\begin{eqnarray}
K_{xx}=K_{yy}=\pi^2 (2\pi)^3 m \frac{1}{w}\int_{0}^{+\infty} k^3  \hat{u}(k)^2 \, dk\nonumber\\
\times \int_{0}^{\pi}\sin^3\theta \delta(\cos\theta) \, d\theta.
\label{k4}
\end{eqnarray}
With the change of variables $s=\cos\theta$, we obtain
\begin{eqnarray}
K_{xx}=K_{yy}=\pi^2 (2\pi)^3 m \frac{1}{w}\int_{0}^{+\infty} k^3  \hat{u}(k)^2 \, dk\nonumber\\
\times \int_{-1}^{+1} (1-s^2) \delta(s) \, ds,
\label{k5}
\end{eqnarray}
so that, finally,
\begin{eqnarray}
K_{xx}=K_{yy}=8 \pi^5  m \frac{1}{w}\int_{0}^{+\infty} k^3  \hat{u}(k)^2 \, dk.
\label{k6}
\end{eqnarray}
On the other hand,
\begin{eqnarray}
K^{zz}=2\pi^2 (2\pi)^3 m \frac{1}{w} \int_{0}^{+\infty} k^3 \hat{u}(k)^2\, dk \nonumber\\ \times\int_{0}^{\pi}\sin\theta\cos^{2}\theta \delta(\cos\theta)  \, d\theta=0.
\label{k7}
\end{eqnarray}
In conclusion, we obtain
\begin{eqnarray}
K^{\mu\nu}=A\frac{w^2\delta^{\mu\nu}-w^{\mu} w^{\nu}}{w^3},
\label{k8}
\end{eqnarray}
with
\begin{eqnarray}
A=8 \pi^5  m \int_{0}^{+\infty} k^3  \hat{u}(k)^2 \, dk.
\label{k9}
\end{eqnarray}
Using Eq. (\ref{ex1}), this leads to Eq. (\ref{vl3}).

\section{Another derivation of the Landau equation}
\label{sec_bb}

For an infinite homogeneous system, the distribution function and the two-body correlation function can be written as $f(0)=f({\bf v},t)$ and $g(0,1)=g({\bf r}-{\bf r}_1,{\bf v},{\bf v}_1,t)$. In that case,  Eqs. (\ref{gen1}) and (\ref{gen2}) become
\begin{eqnarray}
\frac{\partial f}{\partial t}({\bf v},t)=m^2\frac{\partial}{\partial {\bf v}}\cdot \int \frac{\partial u}{\partial {\bf x}} g({\bf x},{\bf v},{\bf v}_1,t)\, d{\bf x}d{\bf v}_1,
\label{ano1}
\end{eqnarray}
\begin{eqnarray}
\label{ano2} \frac{\partial g}{\partial t}({\bf x},{\bf v},{\bf v}_1,t)+{\bf w}\cdot \frac{\partial g}{\partial {\bf x}}({\bf x},{\bf v},{\bf v}_1,t)\nonumber\\
=\frac{\partial u}{\partial {\bf x}}\cdot \left (\frac{\partial}{\partial {\bf v}}-\frac{\partial}{\partial {\bf v}_1}\right )f({\bf v},t)\frac{f}{m}({\bf v}_1,t),
\end{eqnarray}
where we have defined ${\bf x}={\bf r}-{\bf r}_1$ and  ${\bf w}={\bf v}-{\bf v}_1$. 
Using the Bogoliubov ansatz, we shall treat the distribution function $f$ as a constant, and determine the asymptotic value $g({\bf x},{\bf v},{\bf v}_1,+\infty)$ of the correlation function.  Introducing the Fourier transforms of the potential of interaction and of the correlation function, Eq. (\ref{ano1}) can be replaced by
\begin{eqnarray}
\frac{\partial f}{\partial t}=(2\pi)^3 m^2\frac{\partial}{\partial {\bf v}}\cdot \int {\bf k} \hat{u}(k) {\rm Im}\left\lbrack \hat{g}({\bf k},{\bf v},{\bf v}_1,+\infty)\right\rbrack \, d{\bf k}d{\bf v}_1,\nonumber\\
\label{ano3}
\end{eqnarray}
where we have used the reality condition $\hat{g}(-{\bf k})=\hat{g}({\bf k})^*$. On the other hand, taking the Laplace-Fourier transform of Eq. (\ref{ano2}) and assuming that no correlation is present initially (if there are initial correlations, their effect becomes rapidly negligible), we get
\begin{eqnarray}
\tilde{g}({\bf k},{\bf v},{\bf v}_1,\omega)=i\hat{u}(k)\frac{1}{\omega({\bf k}\cdot {\bf w}-\omega)} \nonumber\\
\times {\bf k}\cdot \left (\frac{\partial}{\partial {\bf v}}-\frac{\partial}{\partial {\bf v}_1}\right )f({\bf v},t)\frac{f}{m}({\bf v}_1,t).
\label{ano4}
\end{eqnarray}
Taking the inverse Laplace transform of Eq. (\ref{ano4}), and using the residue theorem, we find that the asymptotic value $t\rightarrow +\infty$ of the correlation function, determined by the pole $\omega=0$, is
\begin{eqnarray}
\hat{g}({\bf k},{\bf v},{\bf v}_1,+\infty)=\hat{u}(k)\frac{1}{{\bf k}\cdot {\bf w}-i0^+} \nonumber\\
\times{\bf k}\cdot \left (\frac{\partial}{\partial {\bf v}}-\frac{\partial}{\partial {\bf v}_1}\right )f({\bf v},t)\frac{f}{m}({\bf v}_1,t).
\label{ano5}
\end{eqnarray}
Using the Plemelj formula 
\begin{eqnarray}
\frac{1}{x\pm i0^+}={\cal P}\left (\frac{1}{x}\right )\mp i\pi\delta(x),
\label{ano6}
\end{eqnarray}
we get
\begin{eqnarray}
{\rm Im}\left\lbrack\hat{g}({\bf k},{\bf v},{\bf v}_1,+\infty)\right\rbrack=\pi \hat{u}(k)\delta({\bf k}\cdot {\bf w}) \nonumber\\
\times {\bf k}\cdot \left (\frac{\partial}{\partial {\bf v}}-\frac{\partial}{\partial {\bf v}_1}\right )f({\bf v},t)\frac{f}{m}({\bf v}_1,t).
\label{ano7}
\end{eqnarray}
Substituting Eq. (\ref{ano7}) in Eq. (\ref{ano3}), we obtain the Landau equation (\ref{vl2}).

\section{Lenard-Balescu equation for homogeneous stellar systems}
\label{sec_lb}

If we assume that the system is spatially homogeneous (or make the local approximation), and take collective effects into account, the Vlasov-Landau equation (\ref{vl2}) is replaced by the Vlasov-Lenard-Balescu equation
\begin{eqnarray}
\frac{\partial f}{\partial t}+{\bf v}\cdot
{\partial f\over\partial {\bf r}}+\frac{N-1}{N}\langle {\bf F}\rangle\cdot {\partial
f\over\partial {\bf v}}=\pi (2\pi)^3 m \frac{\partial}{\partial {v}^{\mu}} \int
k^{\mu} k^{\nu}\nonumber\\
\times \delta ({\bf k}\cdot {\bf w})\frac{\hat{u}^2({k})}{|\epsilon({\bf k},{\bf k}\cdot {\bf
v})|^{2}}
 \left (f_1\frac{\partial
f}{\partial v^{\nu}}-f\frac{\partial f_1}{\partial
v_1^{\nu}}\right )\, d{\bf v}_1 d{\bf k},\qquad
\label{lb1}
\end{eqnarray}
where $\epsilon({\bf k},\omega)$ is the dielectric function
\begin{eqnarray}
\epsilon({\bf k},\omega)=1+(2\pi)^{3}\hat{u}({k})\int \frac{{\bf k}\cdot \frac{\partial f}{\partial {\bf v}}}{\omega-{\bf k}\cdot {\bf v}}\, d{\bf v}. \label{lb1b}
\end{eqnarray}
The Landau equation is recovered by taking $|\epsilon({\bf k},{\bf
k}\cdot {\bf v})|^{2}=1$. The Lenard-Balescu equation generalizes the Landau equation by replacing the bare potential of interaction $\hat{u}(k)$ by a ``dressed'' potential of interaction
\begin{eqnarray}
\hat{u}_{dressed}({\bf k},{\bf k}\cdot {\bf
v})=\frac{\hat{u}({k})}{|\epsilon({\bf k},{\bf k}\cdot {\bf
v})|}.
\label{lb2}
\end{eqnarray}
The dielectric function in the denominator takes into account the
dressing of the particles by their polarization cloud. In plasma
physics, this term corresponds to a screening of the interactions. The
Lenard-Balescu equation accounts for {\it dynamical} screening since
the velocity ${\bf v}$ of the particles explicitly appears in the
effective potential. However, for Coulombian interactions, it is a
good approximation to neglect the deformation of the polarization
cloud due to the motion of the particles and use the {\it static}
results on screening due to Debye and H\"uckel (1923). This amounts to
replacing the dynamic dielectric function $|\epsilon({\bf k},{\bf
k}\cdot {\bf v})|$ by the static dielectric function $|\epsilon({\bf
k},0)|$. In this approximation, $\hat{u}_{dressed}({\bf k},{\bf k}\cdot {\bf
v})$ is replaced by the Debye-H\"uckel potential
$(2\pi)^3\hat{u}_{DH}(k)=(4\pi e^2/m^2)/(k^2+k_D^2)$ corresponding to
$u_{DH}(x)=(e^2/m^2)e^{-k_D r}/r$. If we make the same approximation
for stellar systems, we find that $\hat{u}_{dressed}({\bf k},{\bf
k}\cdot {\bf v})$ is replaced by
\begin{eqnarray}
(2\pi)^3\hat{u}_{DH}(k)=-\frac{4\pi G}{k^2-k_J^2},
\label{lb3}
\end{eqnarray}   
corresponding to $u_{DH}(x)=-G\cos(k_J r)/r$. In this approximation, the Vlasov-Lenard-Balescu equation (\ref{lb1}) takes the same form as the Vlasov-Landau Eq. (\ref{vl3}) except that $A$ is now given by $A=2\pi m G^2 Q$ with
\begin{eqnarray}
Q=\int_{1/R}^{k_L}\frac{k^3}{(k^2-k_J^2)^2}\, dk,
\label{lb4}
\end{eqnarray}  
where $R$ is the system's size. We see that $Q$ diverges {\it
algebraically}, as $(\lambda_J-R)^{-1}$, when $R\rightarrow \lambda_J$
instead of yielding a finite Coulomb factor $\ln\Lambda\sim \ln N$
when collective effects are neglected\footnote{In plasma physics, on
the contrary, $Q=\int_{0}^{k_{L}} {k^3}/\lbrack {(k^2+k_D^2)^2}\rbrack
\, dk$ is well-behaved as $k\rightarrow 0$. When collective effects are taken into account, there is no
divergence at large scales and the Debye length appears naturally. In
the dominant approximation $Q\sim \ln\Lambda$ with
$\Lambda=n\lambda_D^3$. }. This naive approach shows that collective
effects tend to increase the diffusion coefficient or tend to reduce
the relaxation time. This is the conclusion reached by Weinberg (1993)
with a more precise approach. However, this approach is not fully
satisfactory since the system is assumed to be spatially homogeneous
and the ordinary Lenard-Balescu equation is used. In that case, the
divergence when $R\rightarrow
\lambda_J$ is a manifestation of the Jeans instability that a
spatially homogeneous self-gravitating system experiences when the
size of the perturbation overcomes the Jeans length. For
inhomogeneous systems, the Jeans instability is suppressed so the
results of Weinberg should be used with caution. They suggest, however, an
increase of the relaxation time for an inhomogeneous stellar system
that is not far from being unstable. Heyvaerts (2010) derived a more
satisfactory Lenard-Balescu equation that is valid for spatially inhomogeneous
self-gravitating systems. This equation does not present any
divergence at large scales. However, this equation is complicated and
it is difficult to measure the importance of collective effects.

\section{Multi-species systems}
\label{sec_ms}

It is straightforward to generalize the kinetic theory of stellar systems for several species. The Vlasov-Landau equation (\ref{vl2}) is replaced by
\begin{eqnarray}
\frac{df^a}{dt}=\pi (2\pi)^3 \frac{\partial}{\partial {v}^{\mu}} \int
k^{\mu} k^{\nu} \delta ({\bf k}\cdot {\bf w})\hat{u}^2({k})
 \nonumber\\
\times \sum_b \left (m_b f_1^b\frac{\partial
f^a}{\partial v^{\nu}}-m_a f^a\frac{\partial f_1^b}{\partial
v_1^{\nu}}\right )\, d{\bf v}_1 d{\bf k},\label{ms1}
\end{eqnarray}
where $f^a({\bf r},{\bf v},t)$ is the distribution function of species $a$ normalized such that $\int f^a\, d{\bf r}d{\bf v}=N_a m_a$. We can use this equation to give a new interpretation of the test particle approach developed in Sec. \ref{sec_tp}. We make three assumptions: (i) We assume that the system is composed of two types of stars, the test stars with mass $m$ and the field stars with mass $m_f$; (ii)  we assume that the number of test stars is much lower than the number of field stars; (iii) we assume that the field stars are in a steady distribution $f({\bf r},{\bf v})$. Because of assumption (ii), the collisions between field stars and test stars are negligible so that the field stars remain in their steady state. The collisions between test stars are also negligible, so they only evolve due to collisions with the field stars. Therefore, if we call $P({\bf r},{\bf v},t)$ the distribution function of the test stars (to have notations similar to those of Sec. \ref{sec_tp} with, however, a different interpretation), its evolution is given by the Fokker-Planck equation
\begin{eqnarray}
\frac{dP}{dt}=\pi (2\pi)^3 \frac{\partial}{\partial {v}^{\mu}} \int
k^{\mu} k^{\nu} \delta ({\bf k}\cdot {\bf w})\hat{u}^2({k})
 \nonumber\\
\times  \left (m_f f_1\frac{\partial
P}{\partial v^{\nu}}-m P\frac{\partial f_1}{\partial
v_1^{\nu}}\right )\, d{\bf v}_1 d{\bf k}.\label{ms2}
\end{eqnarray}
The diffusion and friction coefficients are given by
\begin{equation}
\label{ms3} D^{\mu\nu}=\pi (2\pi)^{3}m_f\int k^{\mu}k^{\nu}\delta ({\bf k}\cdot {\bf w})\hat{u}({k})^{2} f_1 \, d{\bf v}_{1}d{\bf k},
\end{equation}
\begin{eqnarray}
F_{pol}^{\mu}=\pi (2\pi)^{3}m\int k^{\mu}k^{\nu}\delta ({\bf k}\cdot {\bf w}) \hat{u}({k})^{2}{\partial f_1\over\partial {v}_{1}^{\nu}} \, d{\bf v}_{1}d{\bf k}.
\label{ms4}
\end{eqnarray}
We note that the diffusion is due to the fluctuations of the gravitational force produced by the field stars, while the friction by polarization is due to the perturbation on the distribution of the field stars caused by the test stars. This explains the occurrence of the masses $m_f$ and $m$ in Eqs. (\ref{ms3}) and (\ref{ms4}) respectively.

Using Eq. (\ref{fp7}) and noting that
\begin{eqnarray}
\label{ms5}
\frac{\partial D^{\mu\nu}}{\partial v^\nu}=\frac{m_f}{m}F_{pol}^\mu,
\end{eqnarray}
we get
\begin{equation}
\label{ms6}
{\bf F}_{friction}=\left (1+\frac{m_f}{m}\right ){\bf F}_{pol}.
\end{equation}
If we assume furthermore that $m\gg m_f$, we find that ${\bf
F}_{friction}\simeq {\bf F}_{pol}$. However, in general, the friction
force is different from the friction by polarization. The other
results of Sec. \ref{sec_tp} can be easily generalized to
multi-species systems.

\section{Temporal correlation tensor of the gravitational force}
\label{sec_tcf}

The diffusion of the stars is due to the fluctuations of the
gravitational force. For an infinite homogeneous system (or in the
local approximation), the diffusion tensor can be derived from the
formula (\ref{fpnew3b}) that is well-known in Brownian theory. This
expression involves the temporal auto-correlation tensor of the
gravitational force experienced by a star. It can be written as
\begin{eqnarray}
\label{tcf2}
\langle {F}^{\mu}(t){F}^{\nu}(t-\tau)\rangle=\frac{1}{m}\int d{\bf r}_1d{\bf v}_1 \, F^{\mu}(1\rightarrow 0,t)\nonumber\\
\times F^{\nu}(1\rightarrow 0,t-\tau)f({\bf v}_1).
\end{eqnarray}
Let us first compute the tensor
\begin{eqnarray}
\label{tcf3}
Q^{\mu\nu}=\frac{1}{m}\int  F^{\mu}(1\rightarrow 0,t) F^{\nu}(1\rightarrow 0,t-\tau)\, d{\bf r}_1.
\end{eqnarray}
Proceeding as in Appendix \ref{sec_sw}, we find
\begin{eqnarray}
Q^{\mu\nu}=(2\pi)^3 m \int d{\bf k} \,  k^{\mu} k^{\nu}  e^{i{\bf k}\cdot {\bf w}\tau}\hat{u}({k})^2.\label{tcf4}
\end{eqnarray}
Introducing a system of spherical coordinates with the $z$-axis in the direction of ${\bf w}$, and using Eq. (\ref{ex1}), we obtain after some calculations 
\begin{eqnarray}
Q^{\mu\nu}=2\pi m G^2 \frac{1}{\tau} \frac{w^2\delta^{\mu\nu}-w^{\mu}w^{\nu}}{w^3}.
\label{tcf5}
\end{eqnarray}
According to Eq. (\ref{sw7}), we have
\begin{eqnarray}
K^{\mu\nu}=\int_0^{+\infty} Q^{\mu\nu}\, d\tau.
\label{tcf6}
\end{eqnarray}
Eqs. (\ref{tcf5}) and (\ref{tcf6}) leads to  Eq. (\ref{vl4}) with $\ln\Lambda$ replaced by 
\begin{eqnarray}
\ln\Lambda'=\int_{t_{min}}^{t_{max}}\frac{d\tau}{\tau},
\label{tcf7}
\end{eqnarray}
where $t_{min}$ and $t_{max}$ are appropriate cut-offs. The upper cut-off should be identified with the dynamical time $t_D$. On the other hand, the divergence at short times is due to the inadequacy of our assumption of straight-line trajectories to describe very close encounters. If we take  $t_{min}=\lambda_L/v_m$ and $t_{max}=\lambda_J/v_m$, we find that $\ln\Lambda'=\ln\Lambda$. In Appendix \ref{sec_sw}, we have calculated $K^{\mu\nu}$ by integrating first over time then over space. This yields the logarithmic factor (\ref{vl5}). Here, we have integrated first over space then over time. This yields the logarithmic factor (\ref{tcf7}). As discussed by Lee (1968), these two approaches are essentially equivalent. We remark, however, that the calculations of Sec. \ref{sec_sw} can be performed for arbitrary potentials, while the calculations of this section explicitly use the specific form of the gravitational potential.

According to Eqs. (\ref{tcf2}), (\ref{tcf3}) and (\ref{tcf5}), the force auto-correlation function can be written as
\begin{eqnarray}
\label{tcf8}
\langle {F}^{\mu}(t){F}^{\nu}(t-\tau)\rangle=2\pi m G^2 \frac{1}{\tau}\int d{\bf v}_1 \, \frac{\delta^{\mu\nu}w^2-w^{\mu}w^{\nu}}{w^3} f({\bf v}_1).\nonumber\\
\end{eqnarray}
In particular
\begin{eqnarray}
\label{tcf9}
\langle {\bf F}(t)\cdot {\bf F}(t-\tau)\rangle=4\pi m G^2 \frac{1}{\tau}\int \frac{f({\bf v}_1)}{|{\bf v}-{\bf v}_1|}\, d{\bf v}_1.
\end{eqnarray}
The $\tau^{-1}$ decay of the auto-correlation function of the gravitational force was first derived by Chandrasekhar (1944) with a different method. This result  has also been obtained, and discussed, by Cohen et al. (1950) and Lee (1968). According to Eqs. (\ref{fpnew3b}) and (\ref{tcf8}), the diffusion tensor is given by\footnote{Actually, when the force auto-correlation function decreases as $t^{-1}$, Eq. (\ref{fpnew3b}) is not valid anymore and $\langle (\Delta {\bf v})^2\rangle$ behaves as $t\ln t$ (Lee 1968).}

\begin{eqnarray}
\label{tcf10}
D^{\mu\nu}=2\pi m G^2 \ln\Lambda' \int d{\bf v}_1 \, \frac{\delta^{\mu\nu}w^2-w^{\mu}w^{\nu}}{w^3} f({\bf v}_1).
\end{eqnarray}
This returns Eq. (\ref{rosen1}) with $\ln\Lambda'$ instead of $\ln\Lambda$.

When $f({\bf v}_1)$ is the Maxwell distribution (\ref{e1}), we can compute the force auto-correlation tensor from Eq. (\ref{tcf8}) by using the Rosenbluth potentials as in Section \ref{sec_rosen}. Alternatively, combining Eqs. (\ref{tcf2}) and (\ref{tcf4}), we have
\begin{eqnarray}
\label{tcf11}
\langle {F}^{\mu}(t){F}^{\nu}(t-\tau)\rangle=(2\pi)^{6} m \int d{\bf k} \,  k^{\mu} k^{\nu}  e^{i{\bf k}\cdot {\bf v}\tau}\hat{u}({k})^2 \hat{f}({\bf k}\tau),\nonumber\\
\end{eqnarray}
where $\hat{f}$ is the three-dimensional Fourier transform of the distribution function. For the Maxwell distribution (\ref{e1}), we obtain
\begin{eqnarray}
\langle {F}^{\mu}(t){F}^{\nu}(t-\tau)\rangle&=&(2\pi)^{3} m \rho\nonumber\\
 &\times&\int d{\bf k} \,  k^{\mu} k^{\nu}  e^{i{\bf k}\cdot {\bf v}\tau}\hat{u}({k})^2 e^{-\frac{k^2\tau^2}{2\beta m}}.
\label{tcf12}
\end{eqnarray}
Introducing a system of spherical coordinates with the $z$-axis in the direction of ${\bf v}$, and using Eq. (\ref{ex1}), we obtain after some calculations
\begin{eqnarray}
\langle {F}^{\mu}(t){F}^{\nu}(t-\tau)\rangle=\left (\frac{3}{2\pi}\right )^{1/2}\frac{2\rho m G^2}{v_m} \frac{1}{\tau} G^{\mu\nu}\left (\sqrt{\frac{3}{2}}\frac{v}{v_m}\right ),\nonumber\\
\label{tcf13}
\end{eqnarray}
where $G^{\mu\nu}(x)$ is defined in Section \ref{sec_diffusion}. In particular,
\begin{eqnarray}
\langle {\bf F}(t)\cdot {\bf F}(t-\tau)\rangle=\frac{4\pi\rho m G^2}{v \tau} {\rm erf}\left (\sqrt{\frac{3}{2}}\frac{v}{v_m}\right ).
\label{tcf14}
\end{eqnarray}
Integrating Eq. (\ref{tcf13}) over time, we recover the expression (\ref{diff5}) of the diffusion tensor for a Maxwellian distribution with $\ln\Lambda'$ instead of $\ln\Lambda$.

Finally, the auto-correlation tensor of the gravitational field at two different points (at the same time) is
\begin{eqnarray}
\langle {F}^{\mu}({\bf 0}){F}^{\nu}({\bf r})\rangle=2\pi n m^2G^2\frac{r^2 \delta^{\mu\nu}-r^{\mu}r^{\nu}}{r^3}.
\label{tcf15}
\end{eqnarray}

\end{document}